\documentclass[twocolumn, floatfix, showpacs, english, preprintnumbers, amsmath, amssymb, superscriptaddress, aps]{revtex4-1}
\usepackage{latexsym}
\usepackage{epsfig,psfrag}
\usepackage{dcolumn}
\usepackage{bm}
\usepackage{graphicx}
\usepackage{color}
\usepackage{esint}
\usepackage{babel}
\usepackage{amsfonts}
\usepackage{enumerate}
\begin{document}
\title{Majorana Entanglement Bridge} 
\author{Stephan Plugge}
\affiliation{Institut f\"ur Theoretische Physik,
Heinrich-Heine-Universit\"at, D-40225 D\"usseldorf, Germany}
\author{Alex Zazunov}
\affiliation{Institut f\"ur Theoretische Physik,
Heinrich-Heine-Universit\"at, D-40225 D\"usseldorf, Germany}
\author{Pasquale Sodano}
\affiliation{International Institute of Physics, Universidade
Federal do Rio Grande do Norte, 59078-400 Natal-RN, Brazil} 
\affiliation{Departamento de Fisica T{\'e}orica e Experimental, Universidade
Federal do Rio Grande do Norte, 59072-970 Natal-RN, Brazil}
\affiliation{INFN, Sezione di Perugia, Via A. Pascoli, 06123 Perugia, Italy}
\author{Reinhold Egger}
\affiliation{Institut f\"ur Theoretische Physik,
Heinrich-Heine-Universit\"at, D-40225 D\"usseldorf, Germany}
\affiliation{International Institute of Physics, Universidade
Federal do Rio Grande do Norte, 59078-400 Natal-RN, Brazil} 

\date{\today}

\begin{abstract}
We study the concurrence of entanglement between two quantum dots in
contact to Majorana bound states on a floating superconducting island.  
The distance between the Majorana states, the 
charging energy of the island, and the average
island charge are shown to be decisive parameters for the 
efficiency of entanglement generation.  
We find that long-range entanglement with basically 
distance-independent concurrence is
possible over wide parameter regions, 
where the proposed setup realizes a  ``Majorana entanglement bridge''.
We also study the time-dependent concurrence obtained after 
one of the tunnel couplings is suddenly switched on, 
which reveals the timescales for generating entanglement.
Accurate analytical expressions for the 
concurrence are derived both for the static and the time-dependent case.
Our results indicate that entanglement formation in interacting
Majorana devices can be fully understood in terms of an 
interplay of elastic cotunneling (also referred to as ``teleportation'')
and crossed Andreev reflection processes. 
\end{abstract}


\maketitle

\section{Introduction}

Over the past few years, the prospect of realizing 
Majorana bound states (MBSs) in superconducting hybrid devices 
has attracted a lot of attention, see
Refs.~\cite{mbsrev1,mbsrev2,mbsrev3} for reviews. 
 MBSs are predicted to emerge, for instance, as end states in proximitized 
one-dimensional Rashba nanowires in a magnetic Zeeman field, 
where signatures for their existence 
have already been reported in transport experiments 
\cite{exp1,exp2,exp3,exp4,exp5,exp6,exp7}.
For these topologically superconducting (TS) nanowires, MBSs 
are located at the respective ends of the wire, and the overlap between 
their wavefunctions becomes exponentially small for large nanowire length $L$. 
MBSs can then intuitively be thought of 
as spatially separated ``half fermions''.  They are under discussion
 as basic ingredients for ``topological'' qubits in quantum 
information applications, where information is nonlocally 
stored and, thus, should be quite robust against local decoherence processes.
Systems with MBSs are also predicted to allow for
non-Abelian Ising anyon braiding statistics.  
Since Ising anyons do not allow for
 universal quantum computation by themselves, it is essential 
to thoroughly understand the physics of Majorana states
coupled to conventional qubits, where the latter can 
be realized in terms of nanoscale quantum dots
\cite{hassler,sau,flensberg,preskill,bonderson,leijnse1,leijnse2,hyart}.

Majorana bound states are described by self-adjoint 
operators, $\gamma_j=\gamma_j^\dagger$, which  anticommute
with each other,
$\{\gamma_j,\gamma_k\}=\delta_{jk}$, and with all other fermion operators. 
A pair of MBSs effectively forms a single nonlocal fermion with
annihilation operator
\begin{equation}\label{fdef}
f=(\gamma_1+i\gamma_2)/\sqrt{2},
\end{equation}  
where the eigenvalue $n_f=0,1$ of the number operator
\begin{equation}\label{hatnf}
\hat n_f=f^\dagger f=i\gamma_1\gamma_2+1/2
\end{equation}
describes the state of the MBS pair \cite{mbsrev1}.
For large $L$, the hybridization energy between the MBSs forming this pair
 becomes exponentially small,
\begin{equation}\label{epsfL}
\epsilon_f \sim \exp(-L/\xi).
\end{equation}
The length $\xi$ refers to the spatial size of a MBS and is related 
to the superconducting coherence length in the proximitized 
nanowire \cite{mbsrev1}.
For $L\to \infty$, the MBS pair is then equivalent to a single
zero-energy fermion.

\begin{figure}
\centering
\includegraphics[width=7.5cm]{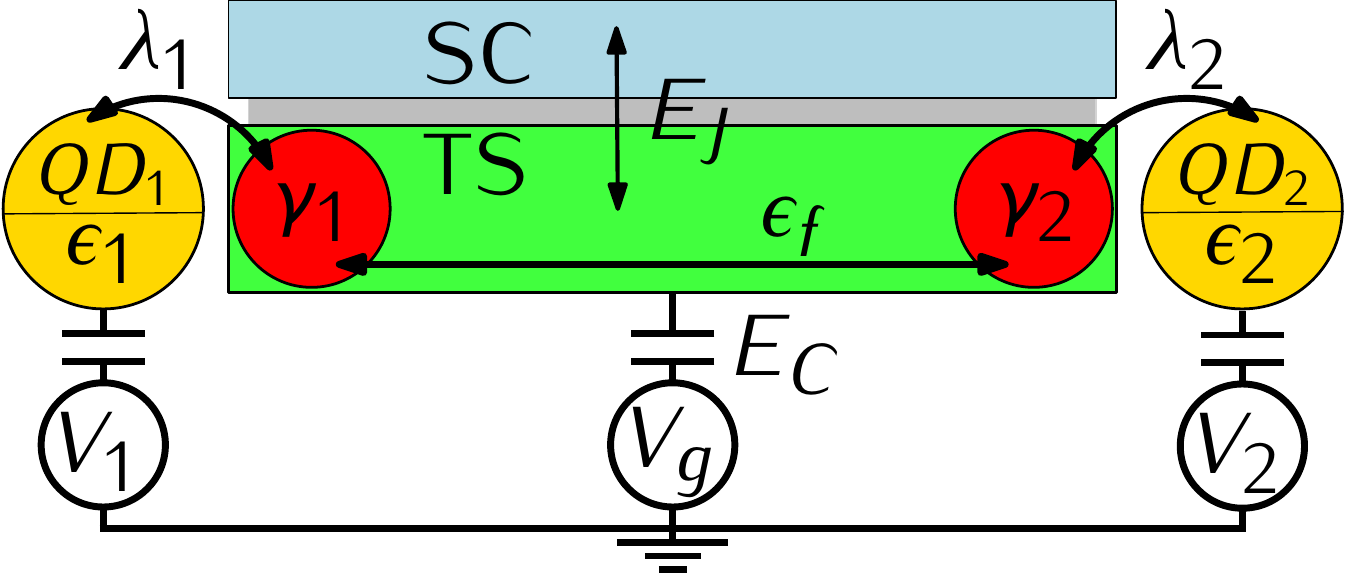}
\caption{\label{fig1} Schematic illustration of the setup.
Large circles correspond to two quantum dots (QD$_1$ and QD$_2$), 
where gate voltages $V_{1,2}$ allow one to vary the energy levels
$\epsilon_{1,2}$.
A Rashba nanowire is deposited on a superconducting island with charging 
energy $E_C$, where the proximity effect in combination
with a Zeeman field induces a TS phase. 
The dimensionless gate parameter $n_g$ (see main text) is 
proportional to a gate voltage $V_g$ regulating the average 
charge on the TS island. 
Small (red) circles represent MBS operators $\gamma_{1}$ and $\gamma_2$, 
where the corresponding wavefunctions are located near the ends of the nanowire 
and the hybridization $\epsilon_f$ 
corresponds to the distance $L$ between both MBSs, see Eq.~\eqref{epsfL}.  
Tunnel couplings $\lambda_1$ and $\lambda_2$ connect the dots to 
the respective MBSs. In addition, the Josephson energy $E_J$ describes the 
coupling to another grounded bulk superconductor (SC).  }
\end{figure}

In this work, we consider the simplest case of a single
MBS pair and study the concurrence of entanglement, $C$,
between two single-level quantum dots \cite{nazarov}
tunnel-coupled to the MBSs, see Fig.~\ref{fig1}
for a schematic illustration of the setup.
Entanglement is a key concept in quantum mechanics and represents
an essential resource for many quantum computation schemes \cite{revent}.  
Two spatially separated systems are said to be entangled when their 
quantum states cannot be described independently.
One may naively expect that because each MBS amounts to half of the 
same fermion state, the MBSs themselves are strongly entangled. 
However, this is a meaningless statement since only the MBS pair has a 
well-defined  state representation, see Eq.~\eqref{fdef}.
On the other hand, the nonlocality of the $f$ fermion level suggests that
the two dots coupled to the MBSs could be fully entangled.  
For a grounded superconducting 
island, where the charging energy $E_C$ is negligible, this question 
has been studied in Refs.~\cite{sodano,bolech,nilsson,tewari,disent},
and one finds that precisely the opposite statement holds true 
in the long-distance limit $\epsilon_f\to 0$, i.e., both dots are 
perfectly disentangled.
This result is consistent with the absence of correlations among 
currents flowing through normal-conducting leads in contact to
 different MBSs \cite{mbsrev1,zazunov3}.

In what follows, we are especially interested in the possibility
of long-range entanglement generation, such that the 
concurrence $C(L)$ does not decay up to very long dot distance $L$ \cite{sodano}.  
We shall determine the detailed conditions for the realization of such
a ``Majorana entanglement bridge'' in the setup shown
in Fig.~\ref{fig1}. In particular, we find that the
charging energy $E_C$ plays an important role in 
allowing for long-range entanglement generation, i.e., 
one should work with a mesoscopic floating (not grounded) superconducting
island. We note in passing that standard notions of long-range entanglement 
refer to an exponential decay, $C(L)\sim \exp(-L/\ell)$, and merely
require that the lengthscale $\ell$ is long \cite{convlrent1,convlrent2}.
In particular, for the topologically trivial limit of our setup, i.e., when the
superconducting island does not host MBSs, such an exponential decay of $C(L)$ 
is expected, see also Ref.~\cite{flensbergnew},
where $\ell$ is equivalent to the superconducting coherence length.
The length-independence of $C(L)$ predicted for the device in Fig.~\ref{fig1} 
holds as long as the charging energy,
which scales approximately as $E_C\sim 1/L$ 
in the TS nanowire geometry, remains the dominant 
energy scale. In particular, $E_C$ should exceed both the thermal energy $k_B T$ 
and the MBS hybridization scale $\epsilon_f$ in Eq.~\eqref{epsfL}.

Indeed, charging effects can change the physics in a profound
way by coupling MBSs to the dynamics of the Cooper pair condensate on the
island.  In Ref.~\cite{fu2010}, Fu studied the transmission through 
a setup as in Fig.~\ref{fig1} but with normal-conducting leads
replacing the dots. For large $E_C$ and when the gate voltage
is tuned to a Coulomb blockade conductance peak,
i.e., the dimensionless parameter $n_g$ regulating the 
average island charge is half-integer,
 only two degenerate island charge states are accessible and
the model effectively maps to a spinless resonant tunneling problem.
 In a slight abuse of terminology, 
the resulting nonlocal phase-coherent transfer of electrons
from one lead to the other has been dubbed ``teleportation'' (TP)
by Fu \cite{fu2010}, see also Refs.~\cite{zazunov1,huetzen}.  
However, it has been argued that entanglement may be achieved 
more efficiently under Coulomb blockade valley conditions, i.e., when
$n_g$ stays close to an integer \cite{wang,budich,foot2}.  
As we discuss in detail below, entanglement is then a 
consequence of elastic cotunneling \cite{nazarov} processes.
In our setup, elastic cotunneling has a strongly nonlocal character 
due to the underlying MBS realization of the $f$ fermion, and therefore
this process has also been referred to as ``teleportation''
\cite{wang,budich}, see Sec.~\ref{sec3a} below. 
When more than two MBSs are present on the superconducting
island, a nonperturbative version of the TP mechanism triggers the 
so-called topological Kondo effect
\cite{beri1,egger1,beri2,altland,zazunov2}. However, in the present work
we shall focus on just one MBS pair, where topological Kondo physics
is absent.

Besides TP processes, however, it is well-known that 
entanglement can also be generated through 
crossed Andreev reflection (CAR) processes. 
So far, CAR-induced entanglement in a system as shown in Fig.~\ref{fig1}
has only been discussed for the grounded device, where $E_C=0$.
As mentioned above, when also $\epsilon_f=0$, the dots will
be perfectly disentangled, but for finite  $\epsilon_f$ (and hence
not too large $L$), CAR processes generate entanglement both for the case of 
normal-conducting leads \cite{sodano,bolech,nilsson,zazunov3} and for
the dot case at hand \cite{tewari,disent}.  
This entanglement can be probed through the violation
 of Bell inequalities when several MBS pairs are present  
\cite{drummond}. In general terms, CAR refers 
to the splitting of a Cooper pair on the island which 
produces entangled electrons on different dots (or leads)
\cite{thierry,loss}.  
Similarly, in the reverse process, a Cooper pair is created by the 
combined in-tunneling of electrons from different sides.
For normal-conducting leads, local Andreev reflection 
competes with CAR, since both electrons can be created (or annihilated) also 
in the same lead, and hence perfect entanglement is 
not possible \cite{nilsson}.  For the single-level dots studied here, however,
local Andreev reflection is absent and perfect entanglement 
due to CAR processes may be realized.  Below we address the 
fate of CAR-generated entanglement as well as
the interplay of CAR and TP processes when interactions ($E_C\ne 0$)
are important.

In addition, we have also studied the time-dependent concurrence, $C(t)$, found 
after one tunnel coupling is suddently switched on. Such a 
``quench'' can be experimentally implemented by changing the 
 voltage on a finger gate in close proximity to the respective tunnel 
contact, cf.~Ref.~\cite{exp1}.  
By monitoring the  time-dependent concurrence,
one can achieve a better understanding of the timescales on which 
entanglement is built up in such a setting \cite{sodano1}.  
For the setup in Fig.~\ref{fig1} with $E_C=0$, this problem has 
recently been studied in Ref.~\cite{disent}.
We here confirm and analytically explain the observations of 
Ref.~\cite{disent} for the noninteracting case. More importantly,  
we also address the entanglement dynamics for finite charging 
energy, and reveal the underlying timescales governing this 
interacting case. 

Before going into a detailed discussion, let us briefly summarize the 
main results of our work:  (i) We 
consider the full crossover behavior of the concurrence 
all the way from vanishing to large charging energy, (ii) allow
for arbitrary $n_g$, (iii) carefully address issues related to 
fermion parity, (iv) include the 
effects of a Josephson coupling $E_J$ to an additional bulk 
$s$-wave superconductor, (v) address the $L$-dependence of 
the concurrence by including the direct overlap $\epsilon_f$ between the MBSs,
and (vi) study the timescales of entanglement generation after
the sudden change of a tunnel coupling.
Throughout the paper, we provide closed analytical expressions for 
the concurrence valid in different parameter regimes, both 
for the static and the time-dependent case, which are shown to give highly
accurate approximations for our numerically exact results.  Importantly,
our results can be quantitatively interpreted in terms of an
interplay of TP and CAR processes,
thereby providing a comprehensive physical framework to describe 
entanglement in MBS devices.  In contrast to Ref.~\cite{wang}, we 
provide a rather complete analysis of this problem, including both TP and
CAR processes, and providing detailed analytical results for the concurrence.

The structure of this paper is as follows.
In Sec.~\ref{sec2}, we briefly summarize the model for the 
setup in Fig.~\ref{fig1}. For the detailed derivation of the
model, we refer the reader to Refs.~\cite{fu2010,zazunov1,huetzen}.
We then turn to a discussion of the entanglement-generating 
tunneling mechanisms in Sec.~\ref{sec3}.
The concurrence, $C$, provides a quantitative measure
for entanglement of the two dots. We compute $C$ analytically in 
several parameter regimes of interest, and validate the results by
comparing to exact numerical diagonalization in Sec.~\ref{sec4}.
The time-dependent concurrence obtained after a quench of
tunnel couplings will then be studied in Sec.~\ref{sec5}.  
Finally, we offer some concluding remarks in Sec.~\ref{sec6}.
Throughout the paper, we focus on the most interesting 
zero-temperature limit and use units with $\hbar=1$.

\section{Model and definitions}
\label{sec2}

\subsection{Low-energy model}

We start by briefly summarizing the low-energy Hamiltonian, 
$H=H_w+H_J+H_d+H_t$, 
describing the setup shown in Fig.~\ref{fig1}, 
see also Refs.~\cite{fu2010,zazunov1,huetzen}.
$H$ contains (i) the part $H_w$ for the central TS island, 
where a mesoscopic superconductor with charging energy $E_C$
causes proximity-induced pairing in the nanowire, (ii) 
the Josephson term $H_J$ coupling the TS island to a second bulk 
superconductor, (iii) the dot Hamiltonian $H_d$, and (iv) a 
tunneling Hamiltonian $H_t$ connecting the dots and the island.
We consider energy scales well below the proximity-induced superconducting
gap, such that it is justified to neglect quasiparticle excitations of the TS
(see also the discussion in Sec.~\ref{sec6}).
The state of the island is then fully described by specifying 
the integer Cooper pair number $N_c$ on the central island, where the 
number operator $\hat N_c$ is canonically
conjugate to the superconductor's phase, $[\hat N_c,
\varphi]_-=-i$, and  the occupation number $n_f$ for the 
$f$ fermion built from the MBS pair, see Eq.~\eqref{fdef}. 
Including the MBS hybridization (\ref{epsfL}) and a Coulomb charging 
term, with the single-electron charging energy $E_C$ and a 
dimensionless gate-voltage parameter $n_g$, 
the island Hamiltonian is given by \cite{fu2010,zazunov1,huetzen}
\begin{equation} \label{hamc}
H_w= \epsilon_f \left( 
\hat n_f-1/2 \right) + E_C\left( 2\hat N_c+\hat n_f-n_g\right)^2,
\end{equation}
where $\hat n_f=f^\dagger f$.  
The nonlocal $f$ fermion representing the MBS pair thus 
couples to the condensate dynamics through Coulomb charging effects.  
Furthermore, the Josephson coupling $E_J$ to another bulk superconductor 
(which is held at constant phase $\varphi_0=0$ and different from the superconductor
responsible for the proximity-induced pairing in the nanowire) is described by
\begin{equation}\label{hamj}
H_J = -E_J\cos\varphi = -\frac{E_J}{2} \left( e^{i\varphi}+e^{-i\varphi}\right),
\end{equation} 
where the operator $e^{i\varphi}$ ($e^{-i\varphi}$)  raises (lowers) 
the Cooper pair number by one unit, $N_c\to N_c\pm 1$.

Next, each quantum dot in Fig.~\ref{fig1} is 
assumed to be in the Coulomb  blockade regime such that
the two dots can be modeled as single fermion levels at energies
$\epsilon_1$ and $\epsilon_2$, see Ref.~\cite{flensbergnew}. 
These levels can be tuned by electrostatic gates  
($V_{j=1,2}$ in Fig.~\ref{fig1})
and by the Zeeman field which induces the TS phase of
the wire and also breaks spin degeneracy on the dots.
As a consequence, we can use the effectively spinless 
fermion annihilation operator, $d_j$, which is connected by a
tunnel coupling $\lambda_j$ to $\gamma_j$, 
see Eq.~\eqref{hamt} below. 
This tunnel coupling also captures a possible spin dependence
of microscopic transition amplitudes and
can be taken as real-valued positive, see Ref.~\cite{zazunov1}. 
The Hamiltonian describing both dots then reads
\begin{equation}\label{hamd}
H_d= \sum_{j=1,2} \epsilon_j \left( \hat n_j - 1/2 \right),
\end{equation}
where the number operator $\hat n_j = d^\dagger_j d^{}_j$ 
has eigenvalues $n_j=0,1$. Equation \eqref{hamd} represents a pair of
charge qubits that can be entangled through the Majorana island.
While charge qubits are more susceptible to detrimental
noise than spin qubits \cite{nazarov}, it is more transparent to analyze 
the concurrence within the present formulation.  
The extension of the above model to the case of spin qubits, where local qubit
operations are also easier to implement, is rather straightforward
by following the route 
sketched in Ref.~\cite{leijnse1}.  However, this case is considerably
more involved on a technical level, since one then effectively
needs two TS nanowires.  We here discuss the simpler spinless case, but
due to the spatial separation of the dots,
the results reported below represent ``useful entanglement'' \cite{revent} 
that could be exploited in quantum computation schemes.

Finally, we come to the tunneling Hamiltonian $H_t$.
 Note that within our low-energy model, no 
TS quasiparticles are available for single-electron tunneling processes, 
and tunneling therefore has to involve MBSs.
We consider sufficiently large $L$ such that $d_1$ is 
tunnel-coupled only to the MBS described by 
$\gamma_1= (f+f^\dagger)/\sqrt{2}$, 
see Eq.~\eqref{fdef} and Fig.~\ref{fig1}.
Similarly, the MBS coupled to $d_2$ corresponds to $\gamma_2\sim f-f^\dagger$.
It is now crucial to take into account charge
conservation, since a floating mesoscopic superconductor 
cannot simply absorb or emit charge $2e$ without energy cost 
--- this is possible only in the grounded case where $E_C=0$.
This consideration implies that tunneling terms 
$\sim d^\dagger f^\dagger$ must include a factor $e^{-i\varphi}$, 
which annihilates a Cooper pair and thus restores charge balance.
With the gauge choice in Ref.~\cite{zazunov1}, we obtain
\begin{equation}\label{hamt}
H_t= \frac{1}{\sqrt{2}} \sum_{j=1,2} \lambda_j d_j^\dagger \left[
f + (-)^{j-1} e^{-i\varphi} f^\dagger \right] + {\rm H.c.} ,
\end{equation}
where ``normal'' tunneling terms like $d^\dagger f$ 
do not affect the condensate but ``anomalous'' tunneling terms 
such as $d^\dagger e^{-i\varphi} f^\dagger$ change the Cooper pair number
by one unit.  

In what follows, it will be convenient to choose occupation number basis states,
\begin{equation}\label{statelabel}
 |n_1n_2n_f,N_c\rangle= \left (d_1^\dagger\right)^{n_1} 
\left (d_2^\dagger\right)^{n_2}
\left (f^\dagger\right)^{n_f} |000,N_c\rangle,
\end{equation}
to represent the Hamiltonian.  For this basis choice, the quantum numbers 
take the values $n_{1,2,f}=0,1$ and $N_c=-N_{\rm max},\ldots, N_{\rm max}$,
where $N_{\rm max}$ restricts the number of Cooper pairs.
Here $N_{\rm max}$ is taken relative to $n_g/2$, and 
the $N_{\rm max}\to \infty$ limit of interest is rapidly approached 
when $E_C$ is finite.  The basis size is then $2M=8(2N_{\rm max}+1)$,
and $H$ becomes a $2M\times 2M$ matrix.  
Diagonalization of this matrix yields the ground state $|\Psi\rangle$, which 
in turn determines the amount of entanglement between the two dots as 
explained below. 

\subsection{Concurrence}

In order to quantify entanglement, one has to identify a suitable measure.  
For our bipartite case with two single-level quantum dots, a convenient 
and reliable measure is given by the concurrence, even though its 
construction is rather formal and abstract
\cite{revent}.  The concurrence is expressed in terms of the 
reduced density matrix $\rho_{d}$ for the dots, which follows from 
the full density matrix after tracing over the island 
degrees of freedom. In the zero temperature limit, this yields
the Hermitian $4\times 4$ matrix 
\begin{equation}\label{rhoddef}
\rho_{d}=  {\rm Tr}_{n_f,N_c} |\Psi\rangle\langle \Psi|,
\end{equation}
with trace equal to unity.  
The concurrence is then defined by \cite{revent}
\begin{equation}\label{Cdef}
C = {\rm max}\left( 0, \sqrt{\lambda_1}-\sqrt{\lambda_2} - \sqrt{\lambda_3}-
\sqrt{\lambda_4}\right), 
\end{equation}
with the eigenvalues $\lambda_1\ge \lambda_2\ge \lambda_3\ge \lambda_4$ 
of the matrix 
\begin{equation}
G=\rho_d (\sigma_y\otimes\sigma_y) \rho_d^* (\sigma_y\otimes
\sigma_y),
\end{equation}
 where $\rho_d^*$ denotes the complex conjugate of
$\rho_d$ and the $\sigma_y$ Pauli matrices act in the 
respective $|n_j\rangle$ (with $n_j=0,1$) space corresponding to each of 
the two dots.  Equation \eqref{Cdef} implies $0\le C\le 1$, 
with $C=0$ for separable (non-entangled)
states and $C=1$ for maximally entangled states. 
We note that the concurrence defined in Eq.~\eqref{Cdef}
is a gauge-invariant quantity, i.e., we can use the above
gauge choice with real-valued $\lambda_j$ in Eq.~\eqref{hamt}.  

We have also computed
the negativity \cite{revent}, 
which is an alternative measure for entanglement. 
While detailed results for the negativity differ from those for
the concurrence, the physical conclusions are the same.
From now on, we therefore discuss the 
concurrence only.  Numerically, $C$ follows from Eq.~\eqref{Cdef} once
the full ground state of the total system, $|\Psi\rangle$, 
has been determined by diagonalization of the $2M\times 2M$ Hamiltonian matrix. 

Analytical progress is possible when $|\Psi\rangle$ has a 
simple form.  An important example is given by
\begin{equation}\label{redmats}
|\Psi\rangle= c_1 |000,N_1\rangle+ c_2 |011,N_2\rangle + 
c_3 |101,N_3\rangle+c_4 |110,N_4\rangle, 
\end{equation}
with arbitrary Cooper pair numbers $N_j$ and complex coefficients $c_j$ 
subject to normalization, $\sum_{j=1}^4 |c_j|^2=1$.
Calculating $C$ in Eq.~\eqref{Cdef} for this state, 
we find that at least one pair of number states 
in Eq.~\eqref{redmats} must have identical $(n_f,N_c)$ 
in order to produce entanglement. 
There are two such possibilities, (i) $N_1=N_4$ with 
$|c_1c_4|\geq|c_2c_3|$, or (ii) $N_2=N_3$
with $|c_2c_3|\geq|c_1c_4|$.  In both cases, the concurrence is given by 
\begin{equation}\label{concanal}
C= 2\left||c_1c_4|-|c_2c_3|\right|,
\end{equation}
see also Ref.~\cite{wang}.  Suppressing the $(n_f,N_c)$ indices,
maximally entangled states with $C=1$ thus correspond to the Bell states 
\begin{eqnarray}\label{bell}
|\Phi_A\rangle &=& \frac{1}{\sqrt{2}} \left( |01\rangle+e^{i\phi_A}
|10\rangle\right), \\ \nonumber
|\Phi_B\rangle &=& \frac{1}{\sqrt{2}} \left( |00\rangle+e^{i\phi_B}
|11\rangle\right),
\end{eqnarray}
with arbitrary phases $\phi_{A/B}$.  

\subsection{Parity conservation and symmetry relations}

Let us now address symmetries and conserved quantities for our model.  
We first note that the total fermion parity defined as
\begin{equation}\label{totalpar}
{\cal P} = (-1)^{n_1+n_2+n_f}
\end{equation}
is a conserved quantity since $H$ contains no terms 
mixing states with even and odd total electron number. 
The even-parity (odd-parity) sector has ${\cal P}=+1$ (${\cal P}=-1$),
where the parity-resolved ground states, $|\Psi,{\cal P}\rangle$,
follow by diagonalization of a Hamiltonian matrix of size
$M\times M$ only. 
Conservation of ${\cal P}$ will generally be weakly broken in 
concrete physical realizations due to quasiparticles neglected
in our model.  In the presence of parity relaxation, 
the true ground state then corresponds to the $|\Psi,{\cal P}\rangle$
state with lower energy.  However, we here assume that ${\cal P}$
is conserved on all timescales of interest, such that both 
states $|\Psi,{\cal P}\rangle$ are experimentally relevant and 
the concurrence depends on total fermion parity, $C=C_{\cal P}$.  
Although this assumption represents an experimental challenge
due to the inevitable presence of residual quasiparticle poisoning, 
very recent experiments indicate that the corresponding timescales
may reach minutes in similar devices as considered here \cite{kouwenhoven,marcus}.

We now turn to the $n_g$-dependence
of the parity-constrained concurrence, $C_{\cal P} (n_g)$.
Since a shift $n_g\to n_g+2$ can be absorbed by shifting $N_c\to N_c+1$
in the Hamiltonian, see Eq.~\eqref{hamc}, 
all observables have to be periodic in $n_g$ with period $\Delta n_g=2$. 
(In the presence of parity relaxation, the period will in general be reduced to
$\Delta n_g=1$.)
We thus obtain a first symmetry relation for the concurrence,
\begin{equation}\label{symrel1}
C_{\cal P}(n_g)= C_{\cal P}(n_g+2).
\end{equation}
This relation allows us to restrict $n_g$ to the window $0\le n_g<2$
throughout.  

A second relation follows from an electron-hole-like symmetry property
of $H$ which relates the odd- and even-parity concurrences, 
\begin{equation}\label{symrel2}
C_{\cal P} (n_g; \epsilon_1,\epsilon_2,\epsilon_f) 
= C_{-{\cal P}} (1-n_g; -\epsilon_1,-\epsilon_2,-\epsilon_f).
\end{equation}
In order to derive Eq.~\eqref{symrel2}, we first note that 
$H$ is invariant under the replacement 
$\epsilon_{1,2,f} \to -\epsilon_{1,2,f}$ with   
$n_g \to 1-n_g$, accompanied by a ``particle-hole transformation'' $U$ 
that exchanges creation and annihilation operators,
$f\leftrightarrow f^\dagger$ etc., 
and $e^{i\varphi} \to e^{-i\varphi}$ yielding $N_c \to -N_c$.
This symmetry implies that if we have the ground 
state $|\Psi(\epsilon_{1,2,f}, n_g) \rangle$, we obtain another 
ground state from the relation
\begin{equation}
 |\Psi(-\epsilon_{1,2,f}, 1-n_g)\rangle = U |\Psi(\epsilon_{1,2,f}, n_g)\rangle.
\end{equation}
Now the point is that these two ground states have different total
parities. Indeed, the original state has ${\cal P}= (-1)^{n_1+n_2+n_f}$,
while the transformed one has ${\cal P}'= (-1)^{(1-n_1)+(1-n_2)+(1-n_f)} = 
-{\cal P}$.  We thus arrive at Eq.~\eqref{symrel2}.   In
passing we note that for an island hosting more than one pair of MBSs, 
if an odd number of dots is tunnel-coupled to MBSs,
we instead would have ${\cal P}' = {\cal P}$, and 
Eq.~\eqref{symrel2} does not connect different parity sectors anymore.
Moreover, the above arguments also show that the relative sign between
$N_c$ and $n_f$ in Eq.~\eqref{hamc} is irrelevant.

For our island with a single MBS pair, Eq.~\eqref{symrel2} 
states that the concurrence in the odd-parity sector directly follows
from the even-parity result by simply inverting the energy scales
$\epsilon_{1,2,f}$ and letting $n_g\to 1-n_g$.  
From now on, unless noted otherwise, we therefore describe 
results for the even-parity sector ${\cal P}=+1$ only, where
it is sufficient to consider gate-voltage parameters in
the window $0\le n_g<2$.

\section{Entanglement-generating processes}
\label{sec3}

Before turning to detailed results for the concurrence
of the two dots in Fig.~\ref{fig1}, let us first 
discuss the entanglement-generating processes at work in such 
a system,  where the nonlocality of the $f$ fermion built from the MBS pair
turns out to be crucial.  
When searching for maximally entangled states of the two
dots, Eq.~\eqref{bell} suggests two candidate sets for suitable
superpositions of dot states $|n_1 n_2\rangle$,
\begin{equation}\label{setdef}
A =  \{  |10\rangle, |01\rangle \},\qquad 
B =  \{  |00\rangle, |11\rangle \}.
\end{equation}
In this section, we shall focus on the large-$E_C$ case, 
assuming that $n_g$ is not half-integer such that 
the equilibrium island state is uniquely defined.
Hence the Cooper pair number corresponds to $N_c=0$, and  for
${\cal P}=+1$, we have $n_f=1$ ($n_f=0$) for set A (B).  

To realize the maximally entangled Bell states \eqref{bell}, 
(virtual) tunneling processes through the island now have to provide the
necessary coupling between states within a given set.
The resulting superpositions should ideally have equal weight, 
which is possible when the dot energy levels are adjusted to fulfill the 
condition $\epsilon_1=\epsilon_2$ ($\epsilon_1=-\epsilon_2$) for 
states in set A (B).  Perturbation theory in the tunneling amplitudes 
then shows the existence of two different entanglement-generating mechanisms, 
namely elastic cotunneling (i.e., TP) and CAR. 
In fact, our quantitative results will find a natural
 interpretation in terms of these two mechanisms.

\subsection{Teleportation}\label{sec3a}

The TP mechanism refers to the nonlocal transfer of an electron across the 
TS island.  For example, let us consider the state $|10\rangle$, where
an electron resides in the left dot and the right dot is empty.  
We now study how a coupling to the partner state $|01\rangle$
in set A is established by $H_t$.  
After the in-tunneling process from the left dot, the island charge will change 
by one unit. However, this change 
typically comes with an energy cost of order $E_C$ 
and, therefore, is possible only as a virtual process.  The 
corresponding out-tunneling event restores the equilibrium charge state 
again and thereby may transfer an electron to the right dot.  
In effect, this process is similar to elastic cotunneling \cite{nazarov} 
but with two major differences.  First, TP has a highly nonlocal 
character inherited from the nonlocality of the $f$ fermion. 
Second, because of the existence of ``normal'' and ``anomalous'' tunneling 
terms in $H_t$,
there are two different contributions.  
Using the notation in Eq.~(\ref{statelabel}), the first one (denoted by TP$_n$) proceeds 
solely by ``normal'' tunneling ($\sim d^\dagger f,f^\dagger d$), while 
the second contribution (TP$_a$) only employs ``anomalous'' tunneling processes,
\begin{eqnarray}\label{TPproc}
{\rm TP}_n  &:& |101,0\rangle 
\stackrel{\lambda_2}{\rightarrow} |110,0\rangle
\stackrel{\lambda_1}{\rightarrow} |011,0\rangle,\\ \nonumber
{\rm TP}_a &:&  |101,0\rangle \stackrel{\lambda_1}{\rightarrow} 
|000,1\rangle
\stackrel{\lambda_2}{\rightarrow} |011,0\rangle.
\end{eqnarray}
 By virtue of the processes in Eq.~\eqref{TPproc},
the $|10\rangle$ and $|01\rangle$ states are coupled together.  
For $\epsilon_1=\epsilon_2$, both states enter the superposition with 
equal weight. They may then form the maximally entangled Bell state 
$|\Phi_A\rangle$ in Eq.~\eqref{bell}, where TP is responsible for
long-range entanglement.  

Finally, we note that the TP phenomenon
discussed by Fu \cite{fu2010} is a resonant version of the above 
process, which takes place near half-integer values of $n_g$.  We will address
this case in detail in Sec.~\ref{sec4a}, where we show that 
perfect entanglement is not possible under Coulomb blockade peak
conditions.

\subsection{Crossed Andreev reflection}\label{sec3b}

CAR provides a distinct second entanglement-generating mechanism, 
where for $\epsilon_1=-\epsilon_2$, 
the two dot states in set B enter the superposition with 
equal weight and may yield the Bell state $|\Phi_B\rangle$ in Eq.~\eqref{bell}.
Since entangled states have to involve a pair of number
states with matching $(n_f,N_c)$ entries,
see Eq.~\eqref{concanal}, the two states of interest now 
correspond to $|000,0\rangle$ and $|110,0\rangle$. 
Their total particle numbers are different, and hence not only the tunnel couplings $\lambda_{1,2}$, but also a 
finite Josephson coupling $E_J$ to another bulk superconductor is 
needed in order to yield non-zero concurrence.
We now show that this condition arises even though
CAR processes can connect states with different island particle 
number $2N_c+n_f$ already for $E_J = 0$.
Indeed, starting from $|000,0\rangle$, there are
six different CAR transitions corresponding
to all permutations of the three elementary transfer processes 
$\sim\lambda_{1}, \lambda_{2}, E_J$.
As an example, we show two of these CAR sequences,
\begin{eqnarray}
\nonumber
{\rm CAR}_1  & : &  |000,0\rangle \stackrel{\lambda_2}{\rightarrow} 
|011,-1\rangle \stackrel{E_J}{\rightarrow} |011,0\rangle
 \stackrel{\lambda_1}{\rightarrow} |110,0\rangle, \\ \nonumber
{\rm CAR}_2 & : &  |000,0\rangle \stackrel{E_J}{\rightarrow} 
|000,1\rangle \stackrel{\lambda_1}{\rightarrow} |101,0\rangle
 \stackrel{\lambda_2}{\rightarrow} |110,0\rangle, \\&&
\label{CAR}
\end{eqnarray}
where the first (second) step for the CAR$_1$ (CAR$_2$) process
describes anomalous tunneling, the respective third step 
corresponds to normal tunneling, and the second (first) step involves 
the Josephson coupling in Eq.~\eqref{hamj}.  In the latter step,
a Cooper pair is transferred to the TS island. 
It is only because of this step that the two
states $|000,0\rangle$ and $|110,0\rangle$
can be connected.  Finally, we note that 
in our quantitative analysis below,
not only the two processes in Eq.~\eqref{CAR}
but all six CAR sequences will be taken into account.

\section{MBS-mediated entanglement}
\label{sec4}

In this section, we quantitatively discuss the concurrence $C$,
describing the amount of entanglement between the two dots in 
Fig.~\ref{fig1}, by comparing 
analytical expressions in different parameter regimes to the 
corresponding numerically exact results obtained from the  
ground state of the full system, see Eq.~\eqref{Cdef}.
Since the entire behavior of the concurrence in the odd-parity
sector follows from the even-parity results, see Eq.~\eqref{symrel2}, 
we shall only discuss the case ${\cal P}=+1$.  
Furthermore, because of Eq.~\eqref{symrel1}, the gate parameter $n_g$ 
can be taken in the window $0\le n_g<2$.  

\begin{figure}
\centering
\includegraphics[width=6cm]{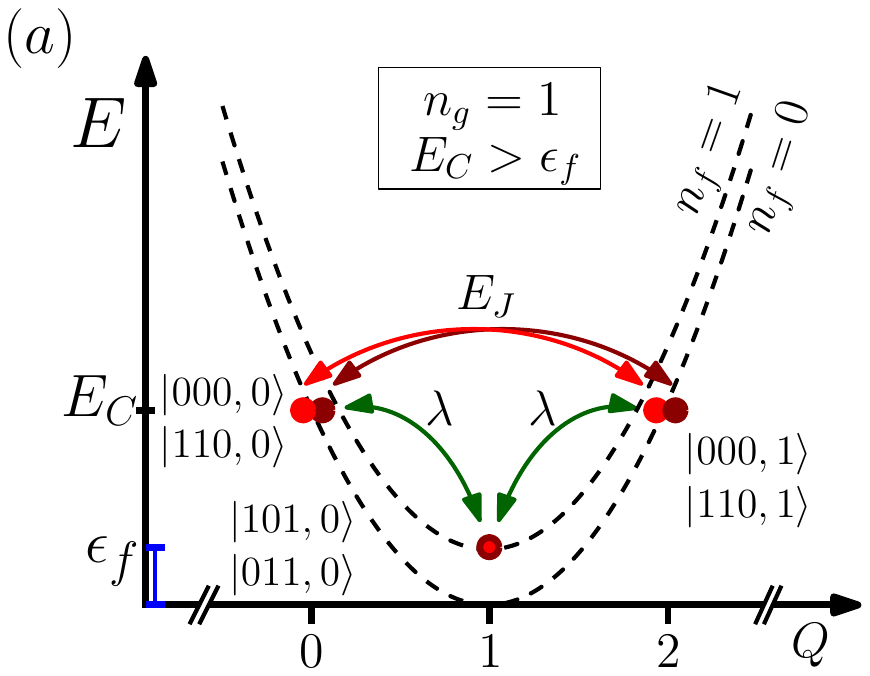}

\vspace*{0.6cm}

\includegraphics[width=6cm]{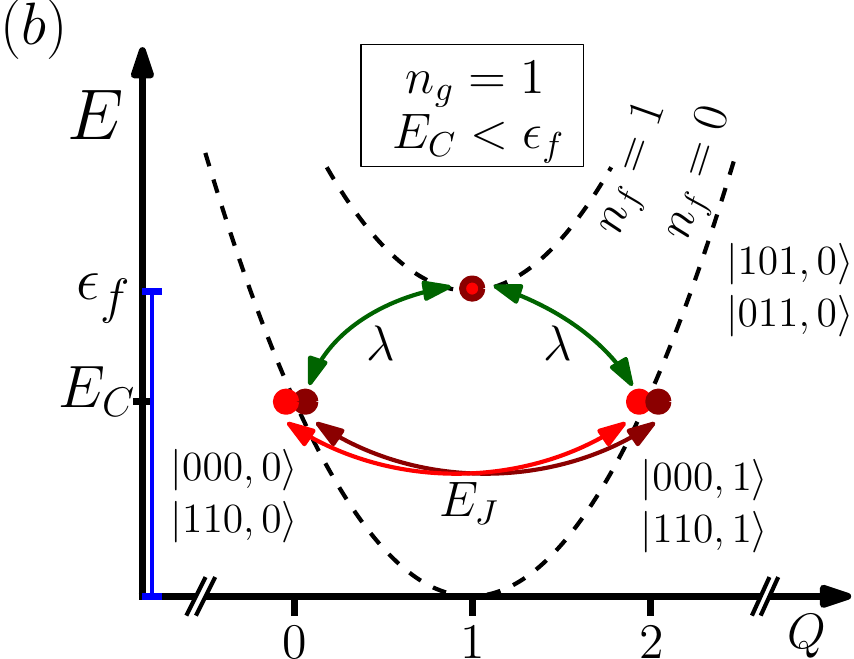}

\vspace*{0.6cm}

\includegraphics[width=6cm]{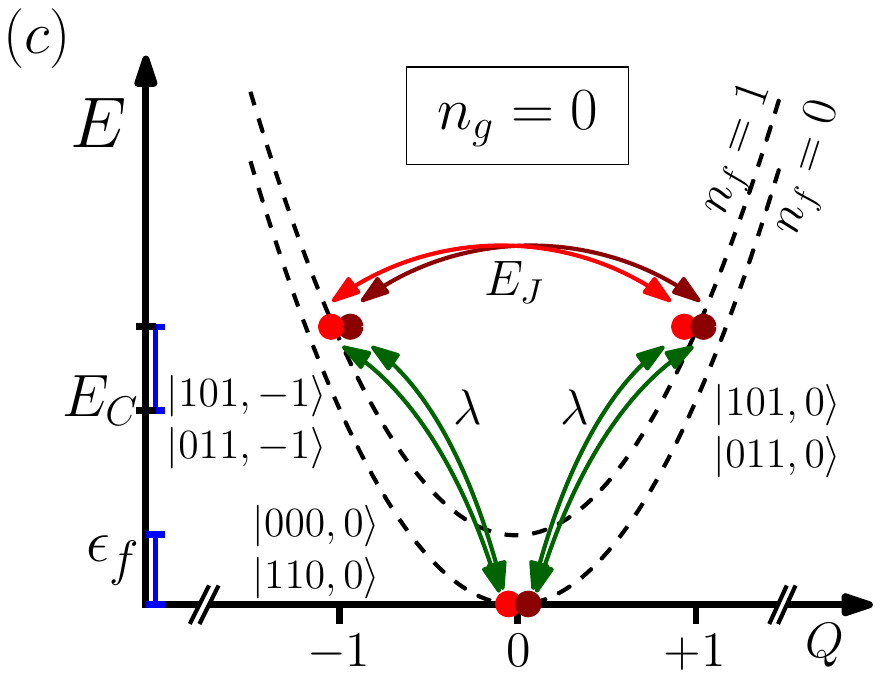}
\caption{\label{fig2} 
Schematic illustration of the energy $E_\nu$, see Eq.~\eqref{Enu}, vs the 
island particle number $Q=2N_c+n_f$ for $\epsilon_{1,2}=0$.  Panel (a)
shows the case $n_g=1$ for $E_C>\epsilon_f$. Panel (b) is also for $n_g=1$ 
but with $E_C<\epsilon_f$, while panel  (c) is for $n_g=0$.
 The respective even-parity states $|\nu\rangle=|n_1n_2n_f,N_c\rangle$  
are also indicated, where tunneling $(\lambda\ne 0)$ and/or 
Josephson $(E_J\ne 0)$ processes cause the shown transitions.  }  
\end{figure}

For convenience, we will employ a compact notation 
for the number states by writing $|\nu\rangle= |n_1n_2n_f,N_c\rangle$ in 
Eq.~\eqref{statelabel}.
Without the Josephson coupling and the tunnel amplitudes,
i.e., for  $E_J=\lambda_{1,2}=0$, 
the eigenenergies of the full system are then given by
\begin{equation}\label{Enu}
E_{\nu} = E_C( Q-n_g)^2+ \sum_{j=1,2,f}\epsilon_j(n_j-1/2),
\end{equation}
with the integer island particle number $Q=2N_c+n_f$.
The dependence of $E_\nu$ on $Q$ 
includes the well-known Coulomb charging energy parabola
\cite{nazarov} and is sketched in Fig.~\ref{fig2}.  
We shall return to Fig.~\ref{fig2} below when discussing
our analytical results for the concurrence.
Moreover, the symbol ${\cal N}$ appearing in some equations below
denotes a normalization constant for the respective state.

Before describing our results for the concurrence in detail, we pause 
to offer some guidance for focused readers.  In Sec.~\ref{sec4a}, 
we analyze the limit of strong Coulomb blockade, where the charging
energy is the dominant energy scale and only a few island charge 
states are accessible.  We derive analytical expressions for the
concurrence $C$ for gate-voltage parameters near the Coulomb
blockade valleys centered around $n_g=1$,
see Eq.~\eqref{concurrencefinal},  and $n_g=0$~mod~2, see
Eq.~\eqref{stephannew} below.  In addition, we also provide closed
results valid near a Coulomb blockade peak,
see Eq.~\eqref{conchalf}, and compare our analytical results to full
 numerical diagonalization.
We then briefly discuss the noninteracting ($E_C=0$) limit in
Sec.~\ref{sec4b}, where the concurrence follows as given in
 Eq.~\eqref{Cnonint}.  Finally, we return to the strong-$E_C$ limit
in Sec.~\ref{sec4c} but now also allow for a strong MBS 
hybridization $\epsilon_f$, i.e., arbitrary length $L$ in Eq.~\eqref{epsfL}.  
However, for simplicity, we only study the case of integer $n_g$ in 
Sec.~\ref{sec4c}.

\subsection{Strong Coulomb blockade regime}\label{sec4a}

We start our quantitative analysis of the concurrence with the 
limit of large $E_C$,
\begin{equation}\label{strongcb}
E_C\gg {\rm max}|\epsilon_{1,2,f}, \lambda_{1,2}, E_J|, 
\end{equation}
where the charging energy dominates over all other energy scales except for the
TS gap, which implicitly sets the ultraviolet cutoff for 
our model. The parameter regime in Eq.~\eqref{strongcb} 
has also been addressed in Refs.~\cite{wang,budich}.
It is convenient to distinguish four different, 
partially overlapping, gate-voltage parameter windows defined by
\begin{equation}\label{gatedef}
n_g=n_{g,0}+\eta/2,\quad n_{g,0}\in\left
 \{ 0, \frac12 , 1, \frac{3}{2} \right\},
\end{equation}
where $|\eta|<1$ parametrizes the deviation from the respective
center value $n_g=n_{g,0}$.  For both half-integer values of
$n_g=n_{g,0}$, the dominant charging energy contribution, given by 
the first term in Eq.~\eqref{Enu}, is precisely degenerate and thus
corresponds to a Coulomb blockade conductance peak \cite{huetzen}.
We present our results for this case later on, but first
turn to the Coulomb blockade valley at $n_{g,0}=1$.

\subsubsection*{Coulomb blockade valley near $n_g=1$}

For $n_{g,0}=1$ in Eq.~\eqref{gatedef}, we observe from Fig.~\ref{fig2}(a) that the ground state has island occupation number $Q=1$ in order to minimize the 
large Coulomb energy, such that $n_f=1$ and $N_c=0$.
For ${\cal P}=+1$, one additional electron then has to occupy 
the dots.  As a consequence, the low-energy sector
is spanned by 
\begin{equation}\label{ostates}
{\cal H}_0 = \{ |101,0\rangle , |011,0\rangle\},
\end{equation} 
with the two states denoted by $|\nu_0\rangle$.
The energetically closest set of
 states, $|\nu_1\rangle\in {\cal H}_1$,
has $Q=0$ or $Q=2$, see Fig.~\ref{fig2}(a), where
the $f$-electron level representing the MBS pair is empty,
$n_f=0$, with $N_c=0$ or $N_c=1$. The two dots are then either both occupied or both empty, resulting in the subspace
\begin{equation}\label{1states}
{\cal H}_1 = \{ |000,0\rangle, |110,0\rangle, |000,1\rangle, |110,1\rangle\}.
\end{equation}
These states are separated from the ${\cal H}_0$ sector by 
an energy of order $E_C$.
This separation of energy scales allows us to derive a low-energy 
Hamiltonian, $\tilde H$, describing the parameter regime in 
Eq.~\eqref{strongcb} for gate-voltage parameters close to $n_{g,0}=1$.
This reduced Hamiltonian is obtained by projecting the full Hamiltonian 
to the subspace ${\cal H}_0$ only.
Using the two basis states $|\nu_0\rangle$ 
in Eq.~\eqref{ostates}, $\tilde H$ corresponds to a 
$2\times 2$ matrix, such that it is straightforward to 
determine the ground state and the concurrence
analytically. 

The above projection is implemented  by disregarding all high-energy 
states beyond $|\nu_1\rangle\in {\cal H}_1$ in Eq.~\eqref{1states} and 
by treating the coupling between the subspaces ${\cal H}_0$ and ${\cal H}_1$
through a Schrieffer-Wolff transformation \cite{altland,lossSW}, 
which here is equivalent to second-order perturbation theory.  
For $E_C>\epsilon_f$, this coupling is due to tunneling ($H_t$) only, 
see Fig.~\ref{fig2}(a), and with the energies $E_\nu$ in Eq.~\eqref{Enu},
we obtain
\begin{eqnarray}\label{ham2nd}
\tilde H & = & \sum_{\nu_0} E_{\nu_0}|\nu_0\rangle\langle \nu_0|+
\frac12 \sum_{\nu_0, \nu'_0} 
 |\nu_0\rangle \langle \nu'_0| \\
\nonumber &\times&\sum_{\nu_1} 
 \left ( 
\frac{1}{E_{\nu_0}-E_{\nu_1}} +\frac{1}{E_{\nu'_0}-E_{\nu_1}} 
\right ) \langle\nu_0|H_t|\nu_1\rangle
\langle\nu_1|H_t|\nu'_0\rangle,
\end{eqnarray}
where the second term describes virtual excursions to the energetically 
higher $|\nu_1\rangle$ states. Using Eq.~\eqref{strongcb}, 
$\epsilon_{1,2,f}$-terms in the respective denominators are small 
against the large charging energy contribution, and up to 
an overall energy shift, we arrive at
\begin{eqnarray}\label{reducedham}
\tilde H&=& \left(\begin{array}{cc}  a & b\\ b & -a \end{array} \right),\\
\nonumber
a &=& \frac{\epsilon_1-\epsilon_2}{2} 
-\frac{\eta(\lambda_1^2-\lambda_2^2)}{2E_C (1-\eta^2)},\\
\nonumber
b &=& \frac{\lambda_1\lambda_2}{E_C(1-\eta^2)},
\end{eqnarray}
where $\eta$ parametrizes $n_g$ around $n_{g,0}=1$, see
Eq.~\eqref{gatedef}.  
The ground state of Eq.~\eqref{reducedham} is
\begin{equation}\label{bell3}
|\Psi\rangle = {\cal N}
\left[ \left(a-\sqrt{a^2+b^2}\right) |101,0\rangle + b |011,0\rangle\right],
\end{equation}
which matches the general form in Eq.~(\ref{redmats}).
The concurrence then directly follows from Eq.~\eqref{concanal},
\begin{equation}\label{concurrencefinal}
C(X_1)= \frac{1}{\sqrt{1+X_1^2}}, \quad X_1= \frac{a}{b},
\end{equation}
stating that $C$ is a universal function of a single parameter $X_1$
which depends on basically all the microscopic parameters 
according to Eq.~\eqref{reducedham}.

For $X_1=0$, Eq.~\eqref{concurrencefinal} yields the ideal value, $C=1$, 
characterizing perfect entanglement.  Indeed, Eq.~\eqref{bell3} then reduces 
to the Bell state $|\Phi_A\rangle$ in Eq.~\eqref{bell}, and  we  conclude
that entanglement is here established by the TP mechanism.  
This condition for perfect entanglement is fulfilled for all
symmetric systems, i.e., for identical dot level energies and tunneling
strengths,
\begin{equation}\label{symmetrics}
\epsilon_1=\epsilon_2=\epsilon,\quad \lambda_1=\lambda_2=\lambda.
\end{equation}
The approximations leading to the reduced Hamiltonian 
(\ref{reducedham}), and hence to the concurrence in
 Eq.~\eqref{concurrencefinal}, eventually break down 
near a Coulomb peak with $|\eta|\to 1$. Nonetheless,
 Eq.~\eqref{concurrencefinal}
shows that $C=1$ can persist over an extended gate-voltage parameter
window around $n_{g,0}=1$.

\subsubsection*{Coulomb blockade valley near $n_g=0$}

In a similar manner, we next study what happens 
around $n_{g,0}=0$. In Fig.~\ref{fig2}(c), we show a subset of the 
relevant states, implementing two of the possible six CAR processes. 
The energy $E_\nu$ is then minimized for $n_f=N_c=0$, and the ${\cal P}=+1$
low-energy sector is spanned by 
\begin{equation}\label{lowen2}
{\cal H}_0^{(n_{g,0}=0)} = \{ |000,0\rangle, |110,0\rangle \},
\end{equation}
where both dot levels are either occupied or empty.
The excited states relevant for CAR processes, cf.~Sec.~\ref{sec3b},
then correspond to
\begin{eqnarray}\label{lowen3}
\notag
{\cal H}_1^{(n_{g,0}=0)} = \{ &|110,-1\rangle,|101,-1\rangle, |011,-1\rangle,\\
&|101,0\rangle, |011,0\rangle, |000,1\rangle \},
\end{eqnarray}
where the first and the last state have island charge $Q =\mp 2$
and thus are highest in energy, $\sim 4 E_C$, while the remaining states
(with $Q = \pm 1$) are separated from the ${\cal H}_0$ sector by an
energy scale $\sim E_C$.

The two states in Eq.~\eqref{lowen2} have different total particle number.
As discussed in Sec.~\ref{sec3b}, see also Fig.~\ref{fig2}(c),
the CAR mechanism connecting those states has to involve a 
finite Josephson coupling $E_J$ 
and therefore constitutes a third-order process. 
In particular, now a transition within the high-energy sector
 in Eq.~\eqref{lowen3} is necessary, and we thus have to include one 
normal tunneling, one anomalous tunneling, and one Josephson process,
i.e., we have to apply $H_t$ twice and $H_J$ once, 
see Eqs.~\eqref{hamt} and \eqref{hamj}.  
The corresponding third-order perturbation theory can again be 
implemented by a Schrieffer-Wolff transformation \cite{lossSW},
which effectively yields $\tilde H \to \tilde H + H^{(3)}$
for the reduced Hamiltonian in Eq.~\eqref{ham2nd}. 
With $\tilde H_t = H_t + H_J$, we find \cite{footthird}
\begin{equation}\label{ham3rd}
H^{(3)}= \sum_{\nu_0,\nu'_0,\nu_1,\nu'_1} 
\frac{\langle\nu_0|\tilde H_t|\nu_1\rangle
\langle\nu_1|\tilde H_t|\nu'_1\rangle 
\langle\nu'_1|\tilde H_t|\nu'_0\rangle}
{(E_{\nu_0}-E_{\nu_1})(E_{\nu_0}-E_{\nu'_1})} 
|\nu_0\rangle \langle \nu'_0|,
\end{equation}
where $|\nu_1\rangle$ and $|\nu'_1\rangle$ are the virtually 
occupied higher-energy states in Eq.~\eqref{lowen3}
and terms $\sim \epsilon_{1,2,f}$ are small against the charging 
energy contributions in the denominator. 
Using the two states in Eq.~\eqref{lowen2} to represent $\tilde H$,
we obtain the same $2\times 2$ matrix as in Eq.~\eqref{reducedham} but 
with the replacements
\begin{eqnarray}\label{reducedham2}
a\to a' &=& -\frac{\epsilon_1+\epsilon_2}{2} +
\frac{\eta(\lambda_1^2+\lambda_2^2)}{2E_C (1-\eta^2)},\\
\nonumber
b\to b' &=& -\frac{3\lambda_1\lambda_2 E_J}{E_C^2(1-\eta^2)(4-\eta^2)},
\end{eqnarray}
where $\eta$ now parametrizes $n_g$ around $n_{g,0}=0$.
We observe that second-order processes only give corrections to the 
state energies ($a'$), while third-order CAR processes produce the
crucial coupling of both states ($b'$) needed for entanglement. 
We note in passing that second-order corrections  $\sim E_J^2$
only give an irrelevant constant energy shift that 
has been dropped in Eq.~\eqref{reducedham2}.
The ground state finally follows as in Eq.~\eqref{bell3}, but
with the basis states in Eq.~\eqref{lowen2} and the replacements in 
Eq.~\eqref{reducedham2}.
Hence the concurrence is again given by Eq.~\eqref{concurrencefinal} but 
with $X_1 \to X_0 = a'/b'$.  

Although these results look very similar, there are crucial differences in the 
entanglement properties when compared to the $n_{g,0}=1$ case.
In particular, although we may have $X_0 = 0$ for a specific gate-voltage
parameter ($\eta$), such that perfect entanglement
is realized, it is not possible to have $C = 1$ over a wide 
parameter range anymore.
In that sense, within the large-$E_C$ regime defined in Eq.~\eqref{strongcb}, 
 entanglement produced by the CAR mechanism 
is less stable than the one caused by TP. To demonstrate  this point,
consider dot energies $\epsilon_1=-\epsilon_2$ 
where CAR processes are most effective, cf.~Sec.~\ref{sec3b},
keeping symmetric tunnel couplings as before, $\lambda_1=\lambda_2=\lambda.$
We then  obtain
\begin{equation}\label{stephannew}
C(X_0)=\frac{1}{\sqrt{1+X_0^2}}, \quad X_0=- \frac{(4-\eta^2)\eta E_C}{3E_J},
\end{equation}
describing a narrow concurrence peak centered around $n_g=0$, 
with height $C(0)=1$ and width $\delta n_g \sim E_J/E_C$.
This concurrence peak is a clear signature of CAR processes  and
can easily be distinguished from the broad and robust $C=1$ plateau 
generated by TP around $n_g=1$. In particular,  the CAR 
entanglement peak quickly disappears with 
increasing $|\epsilon_1 + \epsilon_2|$, 
while it remains stable against variations with 
$\epsilon_1 \approx -\epsilon_2$.  We thus obtain precisely
the opposite behavior as for the TP-generated concurrence plateau.

\subsubsection*{Concurrence near Coulomb blockade peaks}

Next we study the concurrence behavior near half-integer values of $n_g$. 
For simplicity, we discuss the most interesting symmetric case defined in
Eq.~\eqref{symmetrics}. With $n_g$ near $n_{g,0}=1/2$, there are 
four low-energy states corresponding to the subspace
\begin{equation}\label{lowen4}
{\cal H}_0^{(n_{g,0}=1/2)}=\{
|000,0\rangle, |110,0\rangle, |101,0\rangle, |011,0\rangle\}.
\end{equation}
Since these states are already directly coupled by tunneling, the
reduced Hamiltonian $\tilde H$ can simply be obtained by
projecting $H$ to the subspace in Eq.~\eqref{lowen4}, thereby
neglecting all virtual excursions to higher energy states.   
When represented in the basis \eqref{lowen4}, 
up to an irrelevant overall constant, 
$\tilde H$ then takes the form 
\begin{equation}\label{newredham}
\tilde H^{(n_{g,0}=1/2)}=\left( \begin{array}{cccc} \tilde E-2\epsilon 
& 0 & 0 & 0\\ 0 & \tilde E & \lambda/\sqrt{2}
& -\lambda/\sqrt{2} \\ 0 & \lambda/\sqrt{2} & -\tilde E & 0\\
0 & -\lambda/\sqrt{2}& 0 & -\tilde E \end{array}\right),
\end{equation}
with the $n_g$-dependent energy scale
\begin{equation}\label{tildeE}
\tilde E = E_C (n_g-1/2) + (\epsilon-\epsilon_f)/2.
\end{equation}
The ground state of $\tilde H$ is either given by
\begin{eqnarray}\label{newpsi}
|\Psi_1\rangle &=& {\cal N} \Bigl[ \left( 
\sqrt{\tilde E^2+\lambda^2}-\tilde E\right) |110,0\rangle
\\ \nonumber && -\frac{\lambda}{\sqrt{2}}|101,0\rangle+
\frac{\lambda}{\sqrt{2}}|011,0\rangle\Bigr],
\end{eqnarray}
or it corresponds to the separable $C=0$
state $|\Psi_2\rangle=|000,0\rangle$. The respective energies are
\begin{equation}
E_1=-\sqrt{\tilde E^2+\lambda^2},\quad E_2=\tilde E-2\epsilon.
\end{equation}
Since $|\Psi_1\rangle$ is of the form in Eq.~\eqref{redmats}, 
we can again use Eq.~\eqref{concanal} to 
directly read off the concurrence, but
$E_2>E_1$  is necessary to have a finite concurrence.
With the Heaviside step function $\Theta$, 
we thus obtain  the concurrence in analytical form,
\begin{equation}\label{conchalf}
C(\tilde E/\lambda,\epsilon/\lambda) =  \frac{  \Theta(E_2-E_1) }
{1 +\left [ \left(
\sqrt{\tilde E^2+\lambda^2}-\tilde E\right)/\lambda\right]^2  },
\end{equation}
which now depends on two parameters.
In order to obtain a non-zero value in Eq.~\eqref{conchalf}, 
the gate parameter must be above a critical value, $n_g> n_0$, with
\begin{equation}\label{etacrit}
n_0 = \frac12 + \frac{\epsilon_f+\epsilon}{2E_C}-
\frac{\lambda^2}{4\epsilon E_C}.
\end{equation}
This relation is only relevant when $\epsilon>0$,
since the condition $E_2>E_1$ is always met for $\epsilon \leq 0$.
For $\epsilon>0$, the concurrence thus exhibits an abrupt jump 
from $C=0$ to a finite value as $n_g$ increases through $n_0$.

On the other hand, it can happen that at some gate parameter 
$n_g^{(C=1/2)}$ (above $n_0$ for $\epsilon>0$), the 
turnover value $C=1/2$ separating the CAR-dominated regime 
(near the $n_g=0$ valley) from the TP-dominated regime (near $n_g=1$) 
is reached.  Using Eq.~\eqref{conchalf}, this point is at $\tilde E=0$,
 corresponding to 
\begin{equation}
n_g^{(C=1/2)}= \frac12 +\frac{\epsilon_f-\epsilon}{2E_C}.
\end{equation}
Note that this expression holds  in particular for $\epsilon\le 0$. 
We therefore see that the gate parameter $n_{g,c}$ determining the
transition through $C=1/2$ is given by 
\begin{equation} \label{ngc12}
n_{g,c}= {\rm max}\left(n_0, n_g^{(C=1/2)}\right) \simeq
 \frac12 + \frac{\epsilon_f+|\epsilon|}{2E_C},
\end{equation}
where the last expression becomes exact for $\lambda\ll |\epsilon|$.
For $n_0>n_g^{(C=1/2)}$, the concurrence transition is an abrupt jump.
Otherwise, it is a smooth transition, and the 
concurrence jump only happens at a gate parameter $n_g=n_0<n_g^{(C=1/2)}$
where $C$ has already dropped to a small value.  

The actual value for $n_{g,c}$ in Eq.~\eqref{ngc12} 
is close to $n_g=1/2$ for large $E_C$, but it exhibits
a systematic shift towards $n_g = 1$ when $|\epsilon|$  increases.
We also note that as $n_g$ moves from $n_{g,c}$ 
towards $n_g=1$, the concurrence given by Eq.~\eqref{conchalf} approaches
the perfect entanglement regime,
$C\to 1$, as predicted by our expansion around $n_{g}=1$. Consequently, 
our results for $C(n_g)$ obtained within different $n_g$ windows 
smoothly match onto each other.

Finally, a similar analysis can be carried out for $n_g$
near $n_{g,0}=3/2$.  The result is
given by Eq.~\eqref{conchalf} again but with the replacements
$n_g\to 2 -n_g$ and $\epsilon\to -\epsilon$. For $\lambda\alt |\epsilon|$,
the transition corresponding to Eq.~\eqref{ngc12} 
now takes place at the gate parameter  
\begin{equation}\label{ngc32}
 n_{g,c} \simeq \frac32 - \frac{\epsilon_f+|\epsilon|}{2E_C},
\end{equation}
which again shifts towards $n_g = 1$ with increasing $|\epsilon|$.

\subsubsection*{Discussion}

\begin{figure}
\centering
\includegraphics[width=6cm]{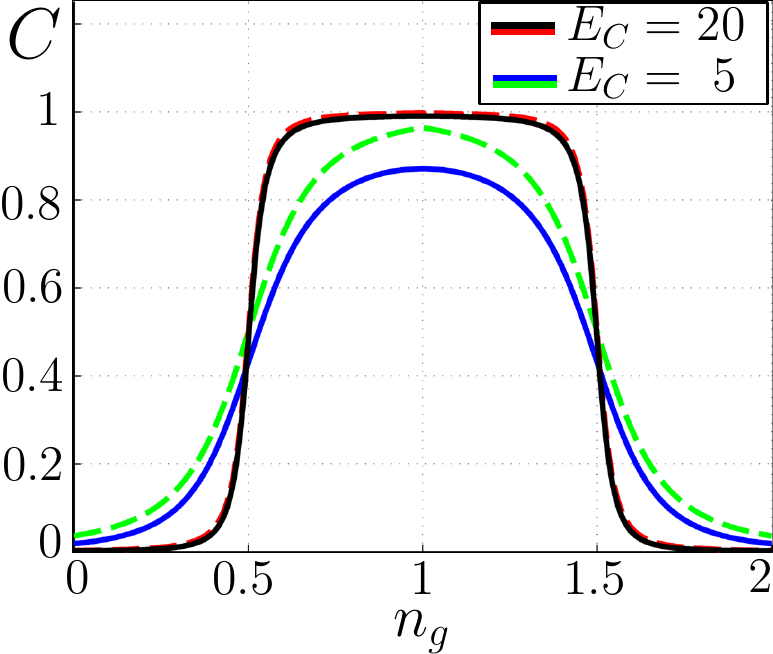}
\caption{\label{fig3}  Concurrence $C$ vs $n_g$ 
for the even-parity sector with $\epsilon_f=E_J=0$. We consider a
symmetric system, see Eq.~\eqref{symmetrics}, with $\epsilon=0$ and $\lambda=1$, for two different values of $E_C$.
Solid curves are obtained by numerical diagonalization of the
full model. Dashed curves represent the analytical result derived in the 
large-$E_C$ limit, see Eqs.~\eqref{concurrencefinal} and \eqref{conchalf}.
With $C(n_g)=C(n_g+2)$, see Eq.~\eqref{symrel1},
we only show one period.}
\end{figure}

With analytical results for the concurrence at our disposal,
we now compare them to numerically exact results.  The 
$n_g$-dependence of the concurrence for $\epsilon_f=E_J=0$ 
is shown in Fig.~\ref{fig3}, where we consider a symmetric system with $\epsilon=0$.  For $E_C=20\lambda$, the analytical curve almost perfectly matches 
the numerically exact result, where we have $C=1$ for practically
the full valley range $1/2<n_g<3/2$, with a
sharp crossover around $n_g=1/2$ to  $C=0$ near even $n_g$. 
For $E_C=5\lambda$, the qualitative behavior is still captured by our 
analytical expressions but there are quantitative differences, as expected when moving away from the strong-$E_C$ limit.  We note in passing that for odd parity, the concurrence behavior in both valleys will effectively be interchanged.

\begin{figure}
\centering
\includegraphics[width=6cm]{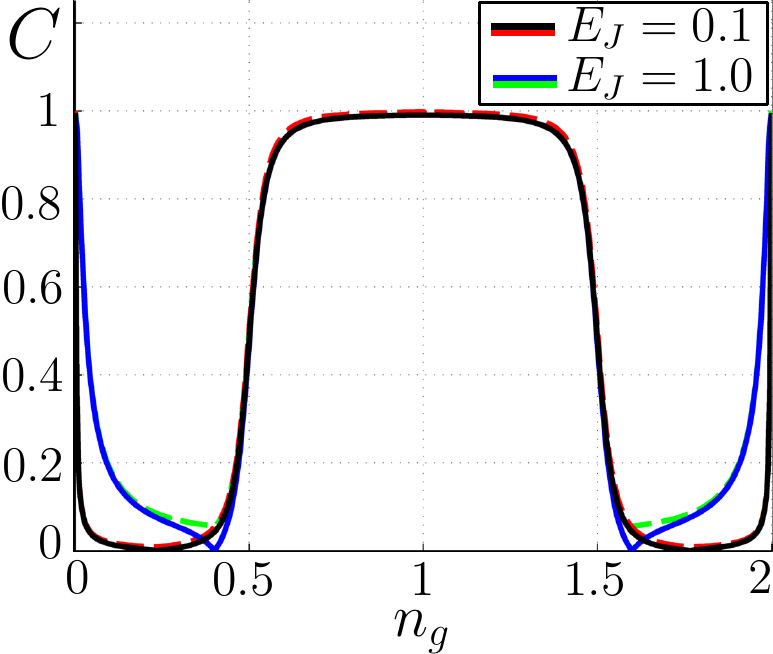}
\caption{\label{fig4} 
Same as Fig.~\ref{fig3} but for finite $E_J$ and $E_C=20$.
The dashed curves again give analytical results, using Eq.~\eqref{stephannew}
near $n_g=0$ and $n_g=2$.  }
\end{figure}

Let us next address the effects of adding a small but finite 
Josephson coupling $E_J$, where we again study a symmetric system 
with $\epsilon=\epsilon_f=0$. The concurrence is shown in Fig.~\ref{fig4}
 for $E_C=20\lambda$.  We first notice that
near  $n_g=1$, the effects of $E_J$ are negligible and the TP-induced 
perfect value $C=1$ 
 is robust against adding a small Josephson coupling. 
However, we now encounter the predicted narrow concurrence peak 
with width $\delta n_g\sim E_J/E_C$ centered at $n_g=0$~mod~2, 
see Eq.~\eqref{stephannew}, which provides
a very characteristic signature of CAR processes.

\begin{figure}
\centering
\includegraphics[width=6cm]{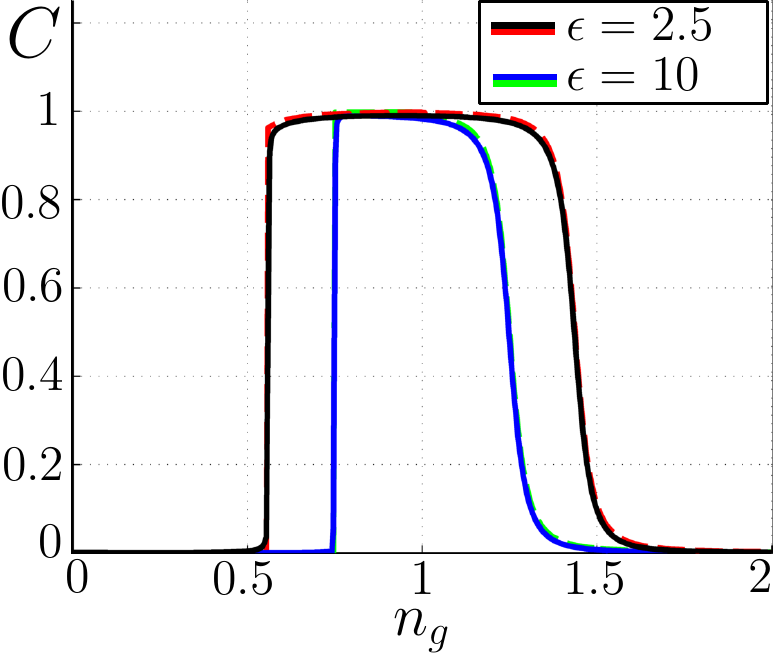}
\caption{\label{fig5}  
Same as Fig.~\ref{fig4} but for finite $\epsilon$, using $E_J=1$.}
\end{figure}

In Figs.~\ref{fig3} and \ref{fig4}, we have considered a symmetric 
system with $\epsilon=\epsilon_f=0$, where the concurrence exhibits 
the additional symmetry $C(n_g)=C(2-n_g)$, and the
turnover between both concurrence regimes is perfectly smooth
and  happens precisely at half-integer $n_g$.  
However, both of these features will not be present in general, and a 
counter-example with finite $\epsilon$ is shown in Fig.~\ref{fig5}.  
In accordance with Eqs.~\eqref{ngc12} and 
\eqref{ngc32}, we find that the turnover points 
between the CAR- and TP-dominated regimes move closer to 
$n_g = 1$ with increasing $\epsilon$.  As expected
from our above discussion, we also confirm that 
for $\epsilon>0$, the crossover in $C(n_g)$
(at $n_g$ slightly above $n_g=1/2$) deforms into an abrupt jump, 
while the corresponding crossover near $n_g=3/2$ stays smooth. 
Hence Fig.~\ref{fig5} clearly shows that $C(n_g)\ne C(2-n_g)$.
We mention in passing that for finite but very 
small $\epsilon$, the narrow entanglement peak seen in Fig.~\ref{fig4}
may survive, but with the center position shifted to finite $n_g$. 
However, this peak is found to quickly disappear with 
increasing $|\epsilon|$, cf.~Sec.~\ref{sec3b}, and is 
altogether absent for the parameters in Fig.~\ref{fig5}.

\subsection{Noninteracting case}\label{sec4b}

Let us now briefly address the noninteracting case, $E_C=0$, where the 
Cooper pair sector decouples from the fermionic sector spanned
 by the $|n_1n_2n_f\rangle$ states.  The $E_C=0$ case
with normal metallic leads instead of our dots was
 studied in Ref.~\cite{nilsson}, while the dot case was also 
considered in Ref.~\cite{disent}.   
An alternative way to reach the noninteracting limit is to 
let $E_J\to \infty$, where all $|N_c\rangle$ states are strongly 
mixed and charge quantization effects are washed out \cite{huetzen}.  
In both limits, $E_C\to 0$ or $E_J\to \infty$, 
one arrives at the same effective Hamiltonian $\tilde H$, where the 
parameters $E_C$, $n_g$ and $E_J$ do not appear anymore.  

In the even-parity sector, $\tilde H$ has a $4\times 4$ matrix 
representation, which can easily be diagonalized to yield the 
ground state $|\Psi\rangle$.  For the symmetric case with $\epsilon=0$, 
see Eq.~\eqref{symmetrics}, and omitting the redundant $N_c$ indices,
we find
\begin{eqnarray}\label{psi0}
|\Psi\rangle &=& {\cal N}\Bigl \{ 
\left(\epsilon_f+\sqrt{\epsilon_f^2+8\lambda^2}\right)
\left[ |000\rangle+|110\rangle \right] \\ \nonumber
&-& \sqrt{8}\lambda \left[ |101\rangle-|011\rangle \right]\Bigr\},
\end{eqnarray}
which is of the form in Eq.~\eqref{redmats}.  
We can therefore infer the concurrence from Eq.~\eqref{concanal},
\begin{equation}\label{Cnonint}
C(X_\infty)= \frac{1}{\sqrt{1+X_\infty^2}},
\quad X_\infty=\sqrt{8} \lambda/\epsilon_f.
\end{equation}
For $\epsilon_f\ll \lambda$, using 
Eq.~\eqref{epsfL}, we thus obtain the conventional exponential decay
of the concurrence with increasing distance between the MBSs, 
$C\sim \epsilon_f/\lambda \sim \exp(-L/\xi)$. 
For $\epsilon_f=0$, the CAR contribution is 
precisely cancelled by the TP contribution corresponding to the 
second pair of states in Eq.~\eqref{psi0}, see also Eq.~\eqref{concanal},
and the concurrence vanishes identically.

On the other hand, for $\epsilon_f\gg \lambda$, we find $C=1$
due to CAR processes, since entanglement is now dominated by 
the first pair of states in Eq.~\eqref{psi0}. 
The noninteracting system thus allows for CAR-mediated 
ideal entanglement as well, but $C=1$ is reached only for rather small
$L$ where  $\epsilon_f\gg\lambda$, see Eq.~\eqref{epsfL}. 
For the realization of long-range entanglement, it is 
therefore essential to work with a floating island where $E_C\ne 0$.

We conclude that the noninteracting case can also be understood 
quantitatively in terms of a competition between TP and CAR processes. 
For even parity, we have seen that CAR processes dominate when a sizeable 
MBS hybridization $\epsilon_f$ is present, which effectively corresponds 
to a rather short distance  $L$ between the MBSs.
We note that for odd parity, the role of TP and CAR processes 
is effectively interchanged, again yielding $C=1$ 
for $\epsilon_f \gg \lambda$. Since we consider single-level dots and
not metallic leads, otherwise competing processes such as normal 
reflection or local Andreev reflection \cite{fu2010,nilsson}
are absent. As a consequence, the entanglement 
found in Eq.~\eqref{Cnonint} is generally more robust.

\subsection{MBS hybridization effects}\label{sec4c}

For small $L$, the MBS hybridization $\epsilon_f$ may exceed the 
charging energy $E_C$ significantly.  We then consider the parameter 
regime \eqref{strongcb} but with arbitrary ratio 
$\epsilon_f/E_C$, i.e., both $\epsilon_f$ and $E_C$ may be large 
compared to all other energy scales in the problem.

\subsubsection*{Gate parameter $n_g=1$}

Let us start with the case $n_g=1$, 
assuming a symmetric setup, see  Eq.~\eqref{symmetrics}. 
As shown before, for $\epsilon_f\ll E_C$, we then have
 $C=1$ due to the TP mechanism. 
Now for finite (but large)
 $\epsilon_f<E_C$,  the energy gap between the 
ground and excited state sectors in Eqs.~\eqref{ostates} and 
\eqref{1states} is given by $\Delta E = E_C -\epsilon_f$, 
see Fig.~\ref{fig2}(a). 
However, as long as the condition in Eq.~\eqref{strongcb}
continues to hold with the replacement $E_C \to \Delta E$,
the situation remains conceptually as before, and for 
$\eta=0$ (i.e., $n_g=1$) and $E_C \to \Delta E$ in 
Eqs.~\eqref{ham2nd}--\eqref{concurrencefinal}, 
we again find $C = 1$.

On the other hand, for $\epsilon_f\gg E_C$, our $E_C=0$ discussion 
suggests that the CAR mechanism generates full entanglement of
 the dots with $C=1$ as well. We therefore now consider 
a finite (large) charging energy with $E_C < \epsilon_f$, where  
the previous ground and excited state sectors are effectively
interchanged,
$\mathcal{H}_0 \leftrightarrow \mathcal{H}_1$ in Eqs.~\eqref{ostates} 
and \eqref{1states}. 
This situation is illustrated in Fig.~\ref{fig2}(b), 
where we can read off  the energy gap separating the two sectors,
$\Delta E'=\epsilon_f-E_C$. With the replacement $E_C\to \Delta E'$,
we then assume that the condition in Eq.~\eqref{strongcb}  holds again.
In particular, the Josephson coupling $E_J$ now directly connects 
the respective states in the low-energy sector such that  
the CAR processes needed for generating entanglement, 
cf.~Sec.~\ref{sec3b}, are already captured by 
second-order perturbation theory, see Eq.~\eqref{ham2nd}.
Considering the case $\epsilon_1 = -\epsilon_2$ and 
$\lambda_1 = \lambda_2 = \lambda$, and diagonalizing the 
remaining $4\times4$ Hamiltonian in the basis \eqref{1states}, 
we obtain the ground state
\begin{eqnarray}\label{psi1}
|\Psi\rangle  &=& {\cal N}\Bigl \{ 
(E_J/2)\left[|000,0\rangle+|110,1\rangle \right] \\ \nonumber
&+&
\left(\Gamma+\sqrt{\Gamma^2+(E_J/2)^2}\right)
\left[|110,0\rangle+|000,1\rangle \right]\Bigr\},
\end{eqnarray}
where the rate
\begin{equation}\label{vtr}
\Gamma = \lambda^2/\Delta E' = \lambda^2/(\epsilon_f-E_C)
\end{equation}
connects $|110,0\rangle$ and $|000,1\rangle$ through the
virtual occupation of excited states.  
Although the concurrence formula \eqref{concanal} is not 
applicable for $|\Psi\rangle$ in Eq.~\eqref{psi1},
$C$ can be computed analytically, with the result
\begin{equation}\label{Clargeef}
C(X_f)= \frac{1}{\sqrt{1+X_f^2}}, \quad X_f = 2\Gamma/E_J.
\end{equation}
In accordance with Sec.~\ref{sec3b}, the concurrence vanishes for 
$X_f\gg 1$, in particular for $E_J\to 0$.

\begin{figure}
\centering
\includegraphics[width=6cm]{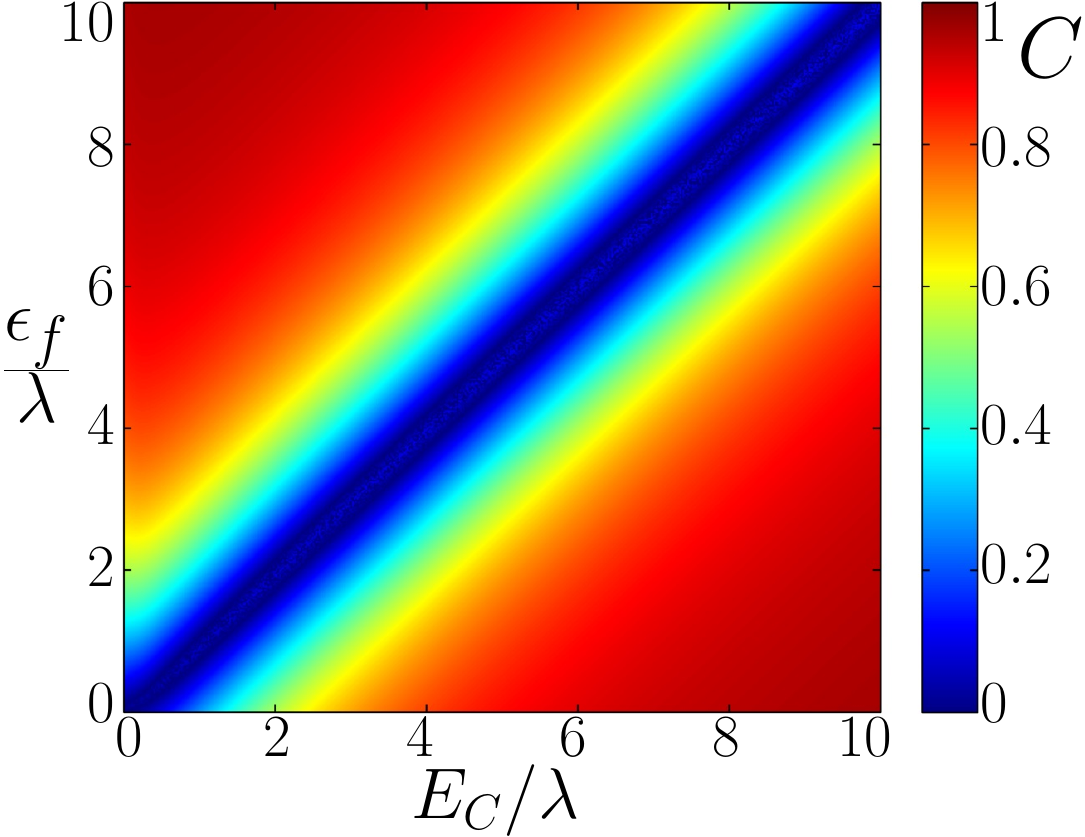}
\caption{\label{fig6}   
Color-scale plot of the concurrence for $n_g=1$ in 
the $\epsilon_f$-$E_C$ plane, taking $E_J=\lambda$ and  a symmetric system 
with $\epsilon=0$, cf.~Eq.~\eqref{symmetrics}.}
\end{figure}

These analytical results are nicely recovered from a 
numerically exact calculation of the concurrence for $n_g=1$.  
We show a color-scale plot of $C$ as a function of $\epsilon_f$ 
and $E_C$ in Fig.~\ref{fig6}, confirming that lines of 
constant concurrence are present when 
$\Delta E = -\Delta E' = E_C -\epsilon_f$ is held constant.
We took $\epsilon_{1,2} = 0$ in Fig.~\ref{fig6}, such that 
both TP and CAR processes can generate stable entanglement,
 see Sec.~\ref{sec3}.  In the crossover regime $\epsilon_f\approx E_C$, 
we find a $C=0$ line right at $\epsilon_f=E_C$, see 
Fig.~\ref{fig6}. This limit is 
correctly captured from the CAR point of view (even though
perturbation theory breaks down),
since then the rate $\Gamma$ in Eq.~\eqref{vtr} becomes large and 
hence $C \to 0$ in Eq.~\eqref{Clargeef}.
Approaching this limit from the TP-dominated side is more
subtle.  The energy diagrams in Fig.~\ref{fig2}(a,b) reveal 
that there are six low-energy states at play for 
$\epsilon_f\approx E_C$. These states are directly coupled via 
$\lambda_{1,2}$ and $E_J$, which can be viewed as a mixing of TP and 
CAR processes. In that case, neither TP nor CAR are effective in 
generating entanglement anymore since the previously excited states may
now be occupied as well.

\subsubsection*{Gate parameter $n_g=0$}

\begin{figure}
\centering
\includegraphics[width=6cm]{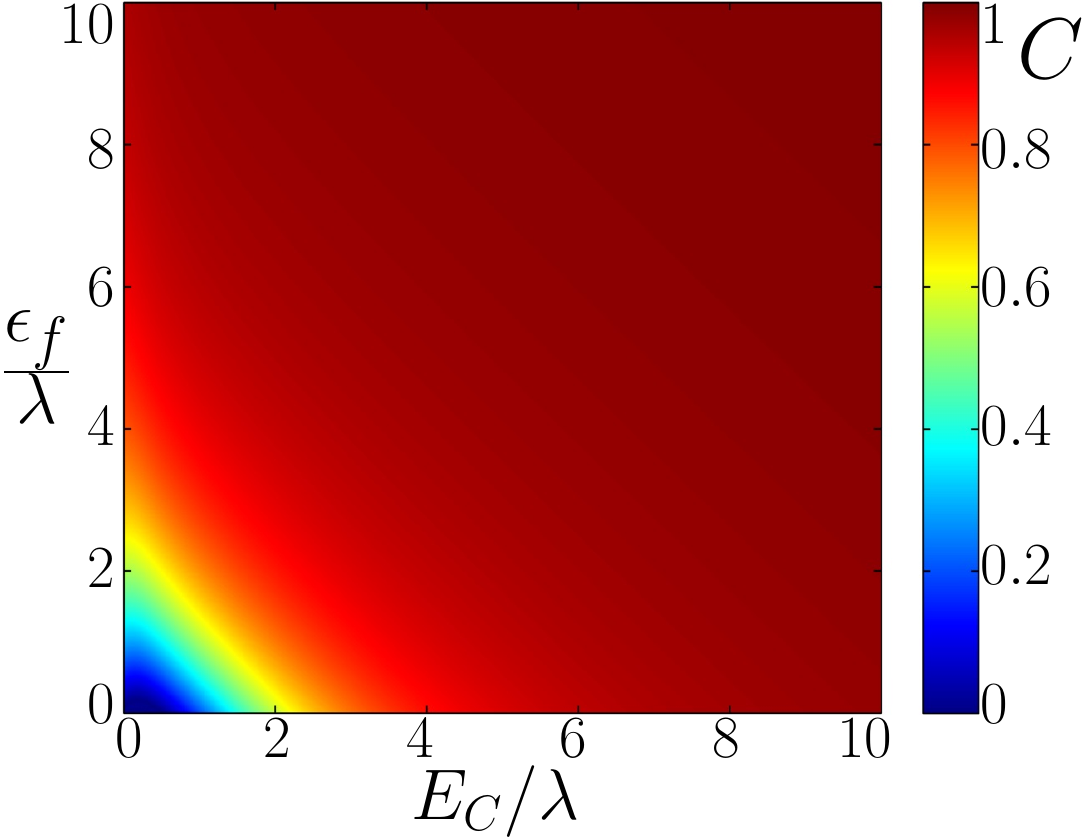}
\caption{\label{fig7}
 Same as Fig.~\ref{fig6} but for $n_g=0$.}
\end{figure}

Finally, we turn to the case $n_g=0$, 
where we consider optimal conditions for the CAR mechanism by 
choosing dot level energies with $\epsilon_1=-\epsilon_2$ and 
symmetric tunnel couplings, $\lambda_1=\lambda_2=\lambda$.
The corresponding energy diagram in Fig.~\ref{fig2}(c) shows 
two out of the six possible CAR processes, see also Sec.~\ref{sec3b}.
With the MBS hybridization $\epsilon_f$,  
the first and the last state in Eq.~\eqref{lowen3} are 
still separated by the energy gap $4E_C$ from the 
ground state sector \eqref{lowen2}, while all other excited
states in Eq.~\eqref{lowen3} have the gap 
 $\Delta E_{0} = E_C + \epsilon_f$.  Depending on the 
ratio $\epsilon_f/E_C$,  different sequences out of the six
possible CAR processes then become important.
Nonetheless, as long as Eq.~\eqref{strongcb} holds for
$E_C\to {\rm min}(4E_C,\Delta E_0)$,
we may effectively apply Eq.~\eqref{reducedham2} again.
We thereby find $\tilde H$ as in Eq.~\eqref{reducedham2} but with 
\begin{equation}\label{reducedham3}
b'\to b'' = -\left( 
\frac{1}{E_C+\epsilon_f} + \frac{1}{2E_C}\right)
\frac{\lambda^2 E_J}{2(E_C+\epsilon_f)} ,
\end{equation}
where we set $\eta = 0$ (i.e., $n_g = 0$). The first term comes
from the two CAR processes with excursions to $Q = \pm 1$
states, see Fig.~\ref{fig2}(c), while the second term describes 
the remaining four CAR processes which also include a virtual occupation 
of $Q=\pm 2$ states.  Regardless of which term dominates, 
throughout the regime 
 $\Delta E_0\gg {\rm max}(\lambda,E_J)$, we observe that
CAR processes dominate the concurrence both 
for $E_C \gg \epsilon_f$ and for $\epsilon_f\gg E_C$.
This fact can be rationalized  by noting that 
the inversion of ground and excited state sectors (discussed 
for $n_g=1$ above) does not take place anymore, since for arbitrary
ratio $\epsilon_f/E_C$, the states $|000,0\rangle$ and $|110,0\rangle$ 
are lowest in energy.
We therefore obtain $C=1$ for almost the entire $\epsilon_f$-$E_C$ plane.
The color-scale plot of the numerically exact solution in 
Fig.~\ref{fig7} confirms this result, with the 
region $\Delta E_0\alt\lambda$ representing the only exception. In 
the latter region, one approaches the perfectly 
disentangled limit $\epsilon_f=E_C=0$, where 
CAR and TP processes interfere destructively.
Finally, we remark that lines of constant concurrence 
are visible in Fig.~\ref{fig7} when
 $b''$ in Eq.~\eqref{reducedham3} is held fixed.

\subsubsection*{Concluding remarks}

Although the above results follow  with the 
replacements $E_C \to \Delta E, \Delta E', \Delta E_0$ in 
Eq.~\eqref{strongcb}, we stress again that $E_C$ 
itself has to remain large,
$E_C > |\epsilon_{1,2},\lambda_{1,2},E_J|$. 
Otherwise the description via charging energy parabolas as in 
Fig.~\ref{fig2} is no longer sufficient and we expect deviations 
from the analytical results. Such deviations can be seen 
for $E_C \alt \lambda = E_J$ in Figs.~\ref{fig6} and \ref{fig7},
where one effectively approaches the noninteracting limit $E_C\to 0$.

\section{Entanglement dynamics}
\label{sec5}

Up to this point, we have studied the stationary case, with 
 time-independent parameters in $H$.  
In this section, we address the entanglement dynamics, $C(t)$, 
observed after a quench of a tunnel coupling
where, say, $\lambda_2$ is suddenly switched on from zero 
to a finite value at time $t=0$.
We assume that at times $t>0$, we then have a symmetric
system with $\epsilon=0$ and $\lambda=1$ in Eq.~\eqref{symmetrics},
and we again consider the even-parity sector, ${\cal P}=+1$.

Studies of the entanglement dynamics can be very useful in revealing 
important timescales of the problem \cite{sodano1,reviewn}.
We will see below that similar to the static case in Sec.~\ref{sec4},
perturbation theory is helpful in elucidating
the underlying physical mechanisms and the relevant timescales
governing $C(t)$.   In particular, our results for $C(t)$ can
again be interpreted in terms of TP and CAR processes.

\subsection{Finite charging energy: $n_g=1$}\label{sec5a}

\begin{figure}
\centering
\includegraphics[width=5cm]{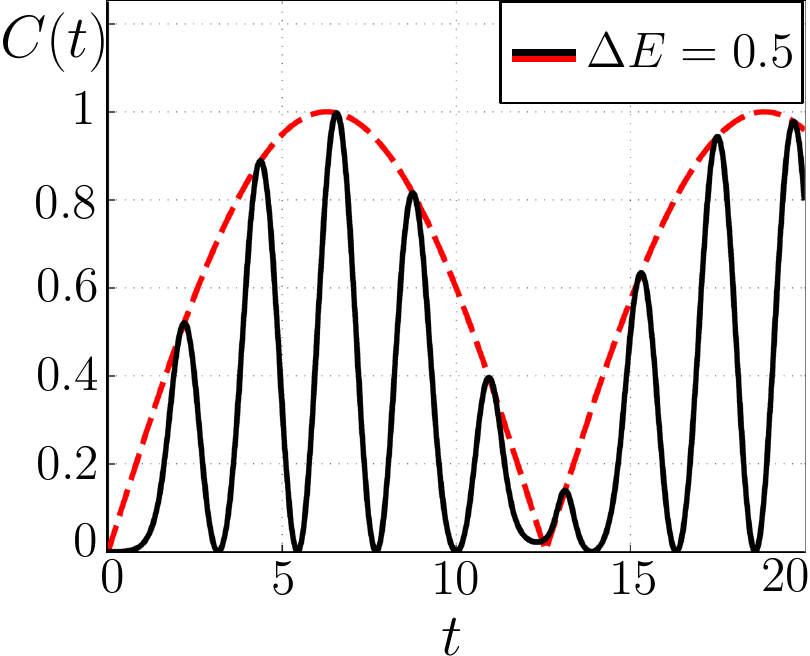}

\vspace*{1cm}

\includegraphics[width=5cm]{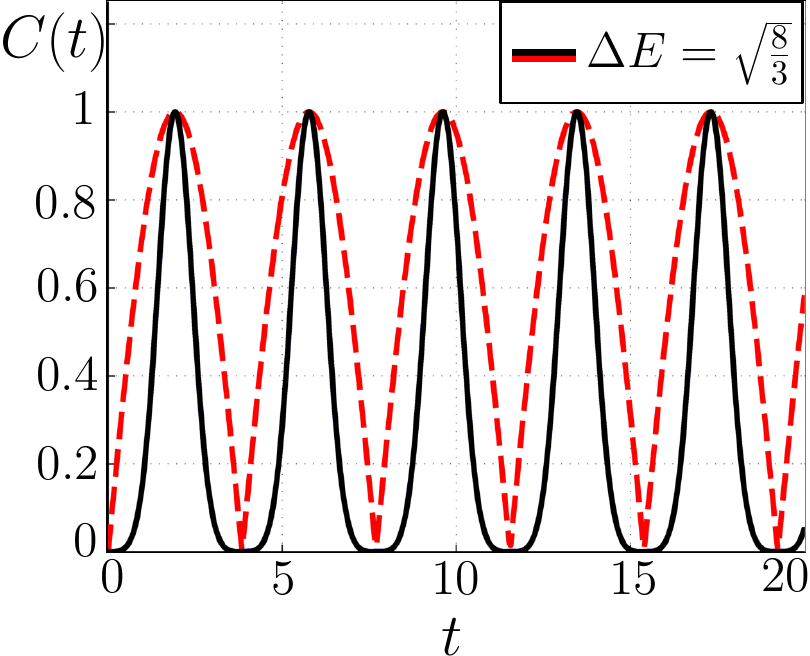}

\vspace*{1cm}

\includegraphics[width=5cm]{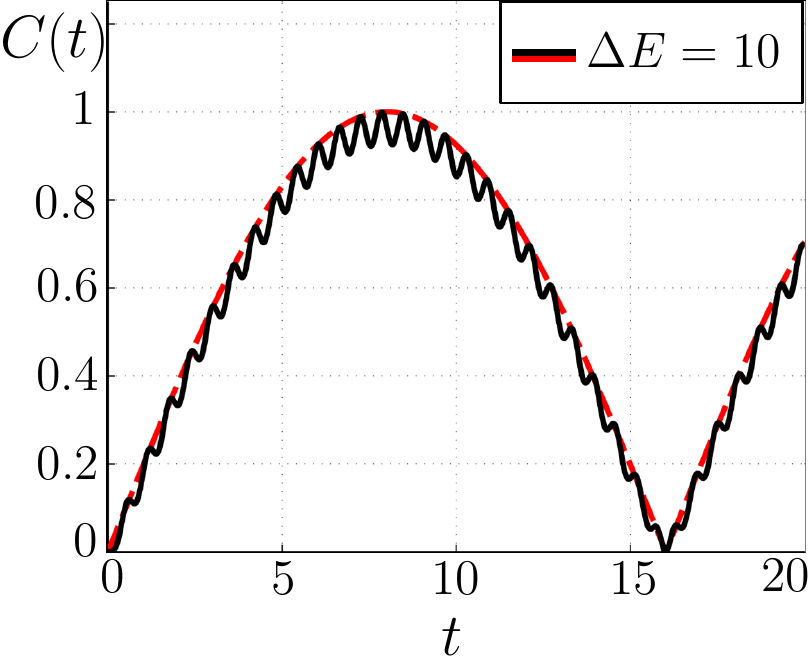}
\caption{\label{fig8}
Time-dependence of the concurrence, $C(t)$, obtained
after a quench of the tunnel coupling,
$\lambda_2=0\to 1$ at $t=0$, for three different values of 
$\Delta E=E_C-\epsilon_f$.  We consider the case of
finite $E_C$ with $n_g=1$, taking $E_J=\epsilon_{1,2}=0$ and 
$\lambda_1=1$.  Exact results (black solid curves) follow from 
Eqs.~\eqref{psit} and \eqref{Ctexact}. 
The envelope functions (red dashed curves) 
in Eq.~\eqref{Cenv} accurately capture the slow part of the dynamics.}
\end{figure}

Let us begin with the case of finite $E_C$ and arbitrary $\epsilon_f$, 
keeping $\Delta E=E_C-\epsilon_f>0$,  see Sec.~\ref{sec4c}.
For simplicity, we put $E_J=0$ and $n_g=1$, where 
we expect that entanglement is mediated by TP processes.
 We note in passing that in the presence of weak parity-breaking 
processes,  e.g., due 
to quasiparticle poisoning, the even total parity (${\cal P}=+1$) sector 
is the true ground state at $n_g=1$.  Suddenly switching on the tunnel 
coupling $\lambda_2=0\to \lambda$ at time $t=0$ 
results in a time-dependence of the concurrence, where our
numerical results for $C(t)$ are shown in Fig.~\ref{fig8}. 
Here we start out from the initial state $|\Psi(t<0)\rangle= |101,0\rangle$,
where the decoupled dot (QD$_2$ in Fig.~\ref{fig1}) is empty
before the quench.  The post-quench state $|\Psi(t>0)\rangle$ 
is of the form
\begin{eqnarray}\label{psit}
|\Psi(t)\rangle &=& c_1(t)|110,0\rangle+c_2(t)|101,0\rangle\\
\nonumber
&+& c_3(t)|011,0\rangle+c_4(t)|000,1\rangle,
\end{eqnarray}
where the time-dependent coefficients $c_j(t)$ are subject to normalization,
 $\sum_j c_j(t) = 1$,  and obey the initial condition $c_j(0) = 
\delta_{j,2}$.  The solution is given by
\begin{eqnarray}\label{Ctexact}
c_{1}(t) &=&-c_4(t)= \frac{-i\lambda}{\sqrt{2} \Omega}
e^{-iE_C t/2} \sin(\Omega t),\\ \nonumber
c_{2}(t) &=& \frac{e^{-i\epsilon_f t/2}}{2}+
\frac{e^{-iE_C t/2}}{2\Omega} \left[ \frac{i\Delta E}{2}\sin(\Omega t)
+\Omega \cos(\Omega t) \right],
\\ c_3(t) &=& e^{-i\epsilon_f t/2}-c_2(t), \nonumber
\end{eqnarray}
where $\Omega=\sqrt{(\Delta E/2)^2+2\lambda^2}$.
The concurrence $C(t)$ then follows by inserting the
coefficients in Eq.~\eqref{Ctexact} into Eq.~\eqref{concanal}. 
Since for $E_J=0$, total particle number 
is conserved, only the given four states in Eq.~\eqref{psit} 
with same total particle number 
as in the initial state are accessible.  For that reason,
Eqs.~\eqref{psit}  and \eqref{Ctexact}
represent the exact solution for arbitrary $\Delta E>0$ 
as long as $E_J=0$.
Indeed, the $C(t)$ results in Fig.~\ref{fig8} are identical to 
those obtained from numerically solving the full Schr\"odinger equation.

To gain physical intuition, it is instructive to
compare these exact (but not very illuminating) results for $C(t)$ to the
 corresponding predictions obtained from the 
reduced $2\times 2$ Hamiltonian $\tilde H$ in Eq.~\eqref{reducedham},
with the replacement $E_C\to \Delta E$ as in Sec.~\ref{sec4c}.
Using the parameters $a$ and $b$ in $\tilde H$, 
only a single oscillation frequency, $\omega_p=2\sqrt{a^2+b^2}$,
appears within that approach, where for our symmetric system,
we find  $\omega_p=2\lambda^2/ \Delta E$.  
The low-energy Hamiltonian $\tilde H$ 
captures only the slow part of the dynamics, i.e., the 
envelope function $C_e(t)$. Moreover, its perturbative derivation 
requires the condition $\Delta E \gg \lambda$.  
On the other hand, for $\Delta E \alt \lambda$, 
we can directly separate the slow and fast variations in the full
analytical expression for $C(t)$ described above,
which gives $C_e(t)$ also for $\Delta E\alt \lambda$.
Combining both cases, we obtain a sinusoidal envelope function,
\begin{equation}\label{Cenv}
C_{e}(t) = \left|\sin(\omega_{e} t)\right|,
\end{equation}
with the frequency
\begin{equation}\label{omgenv}
\omega_e = \begin{cases}
\qquad \Delta E/2,& \Delta E/\lambda < \sqrt{8/3}, \\
\sqrt{(\Delta E/2)^2+2\lambda^2}-\Delta E/2, &\Delta E/\lambda > \sqrt{8/3} .
\end{cases}
\end{equation}
Notice that for $\Delta E/\lambda<\sqrt{8/3}$, the slow part
of the dynamics becomes universal (independent of the tunnel 
coupling $\lambda$).
For $\Delta E/\lambda\approx \sqrt{8/3}$, 
$C(t)$ is governed by two almost resonant frequencies 
and the envelope function in Eq.~\eqref{Cenv} becomes less meaningful.
Figure \ref{fig8} illustrates that
away from this (quite narrow) transition regime, 
Eqs.~\eqref{Cenv} and \eqref{omgenv} accurately 
describe the slow envelopes of all $C(t)$ curves.
For $\Delta E\gg \lambda$, the frequency $\omega_e$ in 
Eq.~\eqref{omgenv} coincides with the perturbative scale $\omega_p$ 
extracted from $\tilde H$.  In this limit, Eq.~\eqref{Cenv}
becomes exact, $C(t)\to C_e(t)$, and the fast oscillations 
in $C(t)$ disappear altogether, cf.~Fig.~\ref{fig8}. 
By comparing the characteristic frequencies appearing in the exact 
solution for $C(t)$ to perturbative estimates valid for $\Delta E
\gg \lambda$, we thus have complemented our picture of the 
system in Fig.~\ref{fig2}.

\subsection{Finite charging energy: $n_g=0$}\label{sec5b}

Such an analysis of the entanglement dynamics becomes
more difficult for other parameters within the large-$E_C$ regime, 
We here briefly consider the case $n_g = 0$ with $E_J > 0$, 
where CAR processes can establish entanglement.  
The Josephson coupling now makes the inclusion 
of higher-energy states with different total particle numbers
necessary, and exact expressions for the envelope curves as 
in Eqs.~\eqref{Cenv} and \eqref{omgenv} are more challenging to obtain.
Nonetheless, analytical estimates are still possible in the perturbative
regime, where we obtain $C_e(t)$ as in Eq.~\eqref{Cenv} but
 with $\omega_{e}\approx \omega_p =2|b''|$, using $b''$ in
 Eq.~\eqref{reducedham3}.
The CAR-induced entanglement dynamics at $n_g=0$ is thus
characterized by a sinusoidal envelope function again but
has the rather long timescale $\sim E_C^2/(\lambda^2E_J)$.

\subsection{Noninteracting case}\label{sec5c}

Another case of interest that allows for an exact solution 
is given by $E_C=0$.
Since the Cooper pair sector decouples, see Sec.~\ref{sec4b}, 
we then have a set of four states describing the full dynamics again.
The concurrence dynamics after a tunnel coupling quench was 
recently addressed for this limit of our setup 
in Ref.~\cite{disent}, where oscillations in $C(t)$ were reported as well.
With $\epsilon_f>0$,  taking the
pre-quench ground state $|\Psi(t<0)\rangle=|110\rangle$, 
and expanding $|\Psi(t)\rangle$ in terms of the four basis states 
available for even parity and $E_C=0$,
the post-quench ($t>0$) state is as in Eq.~\eqref{psit} but with
the initial condition $c_j(0)=\delta_{j,1}$. By 
solving the time-dependent Schr\"odinger equation as detailed in
Ref.~\cite{disent}, $C(t)$ follows again from Eq.~\eqref{concanal}.

 In fact, even though CAR processes are now responsible
for generating entanglement, finding $C(t)$ for $E_C=0$ is
fully equivalent to the TP-mediated case in Sec.~\ref{sec5a} after
replacing $\Delta E = E_C -\epsilon_f \to \epsilon_f$. 
We thus obtain the same concurrence dynamics as in Eqs.~\eqref{Cenv} 
and \eqref{omgenv} after this replacement \cite{footnew}, including the 
results shown in Fig.~\ref{fig8}, where now $\epsilon_f \approx \lambda$ 
separates the perturbative regime ($\epsilon_f \gg \lambda$)  from
the regime where calculation of $C(t)$ needs to account for all four states
($\epsilon_f  \alt \lambda$), see Eqs.~\eqref{psi0} and \eqref{Cnonint}.

\section{Conclusions}
\label{sec6}

In this paper, we have provided a detailed study of entanglement in a 
floating topological superconducting island hosting a pair of MBSs, 
each of which is tunnel-coupled to a single-level dot.  The 
concurrence then provides a convenient measure to quantify entanglement of 
the two dots.  We have shown that this setup offers a robust route towards 
the implementation of a ``Majorana entanglement bridge'' where the concurrence 
does not exhibit any decay with increasing separation of the MBSs.
The only restriction for the separation of the quantum dots then is 
given by requiring a finite charging energy on the island, 
yielding much longer ranges of entanglement than in usual notions 
of exponential decay.  This unique behavior is due to the 
intrinsic nonlocality of the fermion state representing the pair of MBSs. 
The underlying mechanisms for 
entanglement generation have been identified in an intuitive yet 
quantitative manner.  In particular, we have shown that entanglement is 
created by the interplay of teleportation and crossed Andreev reflection 
processes.  With this understanding, we were then able to
 discuss the more complicated entanglement dynamics after
 the quench of a tunnel coupling in terms of simple envelope functions.

We have studied entanglement for the Majorana device 
in Fig.~\ref{fig1} by employing the concurrence as entanglement measure. 
Although the concurrence has a rather formal definition, the
 predicted entanglement features should be observable in 
mesoscopic transport experiments, where the two dots are weakly coupled 
to additional leads as described, e.g., in Ref.~\cite{ensslin}. 
Since current-current correlation functions can detect the violation 
of Bell inequalities \cite{thierry2}, 
signatures of entanglement are expected to
appear in shot noise measurements, see also 
Refs.~\cite{loss2,samuelsson,blaauboer} for related proposals.
In principle also suitably designed conductance measurements 
could probe entanglement \cite{alfredo}.  

Recent experimental work \cite{exp1,exp2,kouwenhoven,marcus} indicates that  
devices corresponding to the model studied here are in close reach. For instance,
charging energies of order $E_C\approx 1$~meV and TS gaps 
$\Delta\approx 0.2$~meV have been reported in Ref.~\cite{marcus}, 
where quantum dots can be formed spontaneously near the TS wire ends 
when contacting them by Au leads \cite{footpr}.  
In order to reach the regime $\Delta > E_C$ assumed in our model, the 
charging energy could be lowered by simply increasing the wire length. 
On the other hand, it may be possible to achieve significant entanglement 
between the dots even when $\Delta < E_C$, where 
fermionic TS quasiparticles should be taken into account. 
When such states are localized in the bulk of the TS such that they do not have 
significant overlap with the Majorana bound states, they have no effect on 
entanglement.  When low-lying quasiparticle states are localized near the 
TS nanowire ends, the physics discussed here may change, but 
entanglement may nonetheless be possible according to our preliminary
analysis, see also Ref.~\cite{flensbergnew}.
Using NbTiN \cite{kouwenhoven} or InAs/Al \cite{marcus} to proximity-induce 
superconductivity in a device similar to the one in Fig.~\ref{fig1},
extremely small poisoning rates have been reported as well.
In any case, a detailed study of quasiparticle effects on entanglement is 
beyond the scope of the present work. 

To conclude, we are confident that future theoretical and experimental studies of
similar setups utilizing such a Majorana entanglement bridge
will benefit from the detailed physical understanding supplied here.

\acknowledgments

We thank S. Albrecht, S. Bose, D. Bruss, E. Eriksson, 
K. Flensberg, H. Kampermann, T. Martin, G. Semenoff, and A.
Levy Yeyati for useful discussions.
R.E.~acknowledges support by the network SPP 1666 of the 
Deutsche Forschungsgemeinschaft (Germany) and by the 
program ``Science Without Borders'' (SWB) of CNPq (Brazil). 
P.S.~thanks the Ministry of Science, Technology, and Innovation of Brazil 
and CNPq for granting a ``Bolsa de Produtividade em Pesquisa'',           
and acknowledges support by the CNPq SWB program 
and from MCTI and UFRN/MEC (Brazil).


\begin{thebibliography}{99}
\bibitem{foot2} In particular, the authors of Ref.~\cite{wang} studied
the large-$E_C$ limit with $\epsilon_f=0$ in a Coulomb valley 
(integer $n_g$) by fixing the number of Cooper pairs, and reported
long-range entanglement in this limit.  By means of a master
equation approach, Ref.~\cite{budich} arrived at similar conclusions 
when the dots are replaced by nanomechanical oscillators.
\bibitem{footthird}
The general expression for $H^{(3)}$ \cite{lossSW} is of symmetrized form
with respect to the initial and final state energies $E_{\nu_0}$
and $E_{\nu'_0}$, cf.~also Eq.~\eqref{ham2nd}.
Since those two energies here coincide, we obtain 
the simpler result in Eq.~\eqref{ham3rd}.
\bibitem{footnew}
At first sight, it might be confusing that even though
 entanglement was generated by TP in Sec.~\ref{sec5a}, 
 CAR processes here yield practically the same results
(with the above replacements).
However, note that for $E_C = 0$, the Cooper pair sector decouples 
completely. Therefore, TP and CAR effectively become equivalent 
virtual transition processes  that only act on different 
sets of states, see Secs.~\ref{sec4b} and \ref{sec4c}.
\end{thebibliography}
\end{document}